%% file: bare_jrnl.tex
\documentclass[journal]{IEEEtran}
\usepackage{xcolor}
\usepackage{amsmath}  
\usepackage{amssymb}    
\usepackage{graphicx}
\usepackage{subcaption}
\usepackage{bm}         
\usepackage{siunitx}    %
\usepackage{cite}        
\usepackage{hyperref}    
\usepackage{url}         
\usepackage{graphicx}
\usepackage{booktabs}
\usepackage{amsmath,amssymb}
\usepackage{siunitx}
\usepackage{float}     
\usepackage{placeins}  
\usepackage{hyperref}
\usepackage{microtype}
\usepackage{stfloats}
\usepackage{multirow}   
\usepackage{booktabs}   
\usepackage{array}      
\usepackage{booktabs}   
\usepackage{multirow}   
\usepackage{array}      
\usepackage{graphicx}      
\usepackage{placeins}      
\usepackage{float}         

\usepackage{amsmath,amssymb,mathtools}
\usepackage{algorithm}
\usepackage[noend]{algpseudocode}
\usepackage{float}        
\usepackage{url}          
\algrenewcommand\algorithmicrequire{\textbf{Input:}}
\algrenewcommand\algorithmicensure{\textbf{Output:}}

\usepackage{graphicx}    
\ifCLASSINFOpdf

\else
 
\fi

\begin{document}

\title{Coordinated Cooling and Compute Management  \\for
 AI  Datacenters }

\author{Nardos~Belay~Abera,~\IEEEmembership{Student Member,~IEEE,} and Yize~Chen,~\IEEEmembership{Member,~IEEE}%
\thanks{N. B. Abera and Y. Chen are with the Department of Electrical and Computer Engineering,
University of Alberta, Edmonton, AB, Canada (e-mail: nbelay@ualberta.ca; yize.chen@ualberta.ca).  This work was supported in part by the Natural Sciences and Engineering
Research Council of Canada, and in part by the Canada First Research Excellence Fund as
part of the University of Alberta’s Future Energy Systems Research Initiative.}%
}

\markboth{IEEE TRANSACTIONS On Cloud Computing}%
{Shell \MakeLowercase{\textit{et al.}}: Bare Demo of IEEEtran.cls for IEEE Journals}
\maketitle

\begin{abstract}
The AI datacenters are currently being deployed on a large scale to support the training and deployment of power-intensive large-language models (LLMs). Extensive amount of computation and cooling required in datacenters increase concerns about the energy use and carbon emissions of AI datacenters. Although  current state-of-the-art has examined the energy efficiency of LLM inference, most prior research focused on optimizing compute-side scheduling without considering thermal objectives or constraints. Since GPU-intensive inference generates substantial heat that can degrade datacenter performance, ignoring thermal effects can increase total energy consumption and reduce the efficiency of LLM serving.  To fill this gap, we profile the characteristics of GPU servers under varying cooling and AI jobs, and develop a joint  cooling and computing modeling approach for AI datacenters. Built upon such workload and thermal dynamics models, a novel hierarchical control framework is proposed to co-optimize computing and thermal management by identifying the  energy-efficient GPU parallelism, frequency (DVFS), and cooling control knobs. Using real Azure inference traces and detailed GPU profiling, our model balances serving latency and thermal constraints in AI datacenters while significantly improving AI datacenters' energy efficiency.
\end{abstract}

\IEEEpeerreviewmaketitle
\section{INTRODUCTION}
\IEEEPARstart{A}{rtificial} intelligence (AI) datacenters are designed to support computing-intensive tasks such as training and serving of the large language model (LLM) and the GenAI model. Due to AI's rapid and widespread adoption, hyper-scale inference clusters are now being deployed to serve trillions of requests per day~\cite{reff1}. To meet the ever-increasing computational requirements of LLMs services, various solutions, such as enhanced datacenter architectures, scheduling algorithms, and accelerators, have been proposed to increase the efficiency of inference~\cite {reff3,reff5,reff6}. Although previous work primarily focused on performance aspects, one critical factor is largely overlooked: the energy perspective of large-scale AI datacenters~\cite{reff7}. This concern is amplified by the substantial economic and environmental impacts of the enormous amounts of energy consumed by modern AI models. In particular, training the 175-billion-parameter GPT-3 would require approximately 1287~MWh of electricity and emit 502~metric tons of $\mathrm{CO}_2$~\cite{reff8}. LLM inference is even gradually dominating the energy consumption landscape. In fact, recent studies project that 80--90\% of total workload and energy consumption of LLMs deployed in production environments arise from inference~\cite{reff9}. Although some previous work has aimed to reduce the energy consumption of LLM inference~\cite{reff11}, these studies explicitly ignore the impact of datacenter thermal behavior and cooling systems~\cite{reff12}.

Such energy-inefficient computing  can be attributed to thermally stressed operating conditions in datacenters, where intense and dense heat generation leads to GPU performance degradation due to thermal throttling~\cite{reff13}. Load and temperature imbalances can also lead to inefficient job scheduling and cooling operations in datacenters. Therefore, neglecting thermal dynamics could result in cooling units reacting only after temperature increases occur~\cite{reff14}. Not only does such operation result in degradation in computing performance, but the energy used for cooling is significant and can be up to 40-50\% of the total energy consumption in AI datacenters~\cite{reff15}.

A major challenge in managing the power and energy profile of AI datacenter is the highly heterogeneous and  time-varying nature of the LLM workload, which varies in context length, output length, and model size. This creates irregular compute demand and rapid heat generation patterns across GPU racks. This heterogeneity breaks the assumption of the traditional datacenter control system  and overwhelms cooling systems, whose slow response leads to temperature overshoots, local hot spots, and inefficient use of cooling energy~\cite{reff16}. Existing datacenter management approaches often decouple computing and
cooling control as separate loops ~\cite{reff17}. Although such methods
improve inference throughput or cooling efficiency in isolation,
they fail to capture the strong interdependence between
computational load, heat generation, and cooling dynamics.
This separation leads to suboptimal energy consumption and
unstable temperature regulation, especially under dynamic
LLM workloads, which ultimately affects the sustainability
and performance of datacenters ~\cite{reff18}.

To address this research gap in the energy management of AI datacenter, this paper proposes a novel hierarchical control framework, which jointly coordinates and controls computing and cooling resources to enable energy-efficient LLM inference. Such a hierarchical framework helps reduce computational complexity and disaggregates multiple control variables, allowing coordinated control across multiple layers. As a foundation for this design, we first  model the power and energy characteristics of the LLM workloads and key cooling control parameters. LLM inference differs fundamentally from conventional CPU-centric datacenter workloads~\cite{reff19,reff20,reff21,reff22}.

\begin{figure}[t]
  \centering
   \includegraphics[
    width=0.98\linewidth,
    clip, trim=2 2 2 2
  ]{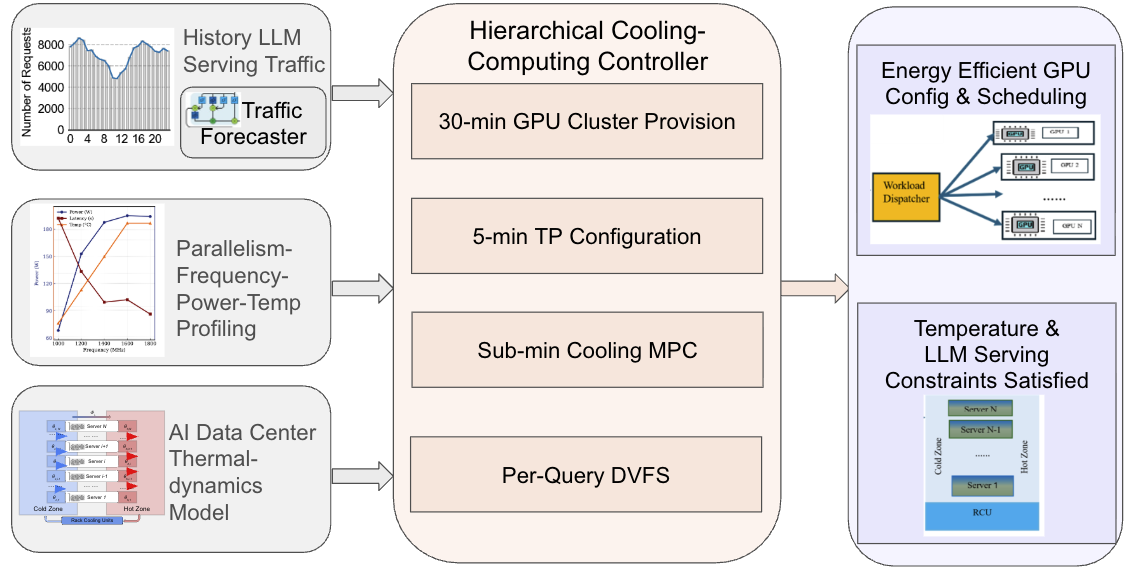}
  \caption{Schematic for proposed hierarchical control of joint cooling–compute for AI datacenters.}
  \label{fig:architecture}
\end{figure}

Taking into account heterogeneous computational behaviors of LLM inference, we develop the AI datacenter heat transfer model, and develop hierarchical control framework operating across multiple levels. At the higher level, we predict the AI workload, and dynamically adjust the number of active servers to maximize utilization and reduce idle power. At the intermediate level, the controller  dynamically changes GPU parallelism, involving the number of GPUs working together on one LLM inference instance, subject to the maximum available GPUs and temperature constraints. The cooling level dynamically adjusts parameters such as the supply air temperature and airflow rate, which vary with respect to workloads. On the lower level, dynamic voltage and frequency scaling (DVFS) is adopted  to adjust the frequencies of GPUs according to the service-level objectives (SLOs) of workloads and thermal conditions. To achieve this, we develop a temperature-aware DVFS mechanism that jointly
considers workload intensity and GPU temperature when selecting the operating
frequency. This enables the controller to reduce energy consumption and regulate temperature without degrading inference performance. This approach also considers the sensitivity of DVFS to the types of work request. Requests with longer prompts are computationally intensive and more sensitive to frequencies, while requests involving short prompts but longer output are memory-bound, which is less sensitive~\cite{reff11}.  The control framework uses detailed profiling data collected from LLM serving
under varying workloads, GPU configurations, and temperatures to capture the
relationships among energy consumption, latency, and temperature, enabling the
controller to determine cooling and computing strategies that satisfy the
specified performance and thermal constraints.  The overall scheme is illustrated in Fig. \ref{fig:architecture}. This paper extends our earlier  paper \cite{reff22b}, which presented a preliminary version of the proposed framework in a short workshop format. The workshop paper primarily introduced the core concept of coordinated cooling and computing management for AI data centers. The current manuscript expands that work through a more comprehensive system model, enhanced optimization and control formulations for LLM inference workloads, and detailed high-performance GPU power and thermal profiling. In addition, the manuscript includes comparisons with state-of-the-art method, as well as robustness and ablation studies that were not presented in the workshop version.

\subsection{Main Contributions}

Building upon LLM serving characteristics and opportunities of co-management of GPU cooling and computing, this work makes the following key contributions:

\begin{itemize}
   \item We propose a novel hierarchical compute--thermal control framework for the AI datacenter  that provides methodology for jointly optimizing computing resources (GPU parallelism and frequency) and cooling resources (supply air temperature and inlet airflow rate). Principled temperature constraints are integrated to guarantee thermal safety while enabling significant energy savings. 
  \item We develop a novel thermal-aware workload dispatch algorithm that schedules jobs based on the capacity of each GPU pool. This algorithm reduces the overhead associated with frequent reconfigurations and ensures smooth adaptation to heterogeneous, time-varying LLM inference workloads. We incorporate an LSTM-based model for forecasting token demand for proactive resource allocation and a DistilBERT-based classifier for classifying each LLM job's length. 
  \item We perform a comprehensive evaluation using Microsoft Azure LLM inference traces in combination with detailed thermal and power profiling of the GPU~\cite{azurellm2024}. Simulation and experimental results demonstrate that the proposed approach reduces both IT-side and cooling-side power consumption while preserving SLOs. The proposed framework is also compared against DynamoLLM \cite{reff11}, adapted TAPAS \cite{reff18} and a single pool baseline to evaluate the benefits of jointly coordinating workload management, GPU DVFS, and cooling control. Overall, the proposed controller achieves 24.2\% computing energy savings and 31.2\% cooling energy savings, and lowers the average GPU temperature by 17.0\%, while maintaining negligible impact on inference latency.
\end{itemize}

\subsection{Article Outline}

The remainder of this paper is organized as follows.
Section~\ref{2} provides background on datacenter cooling systems and the major control knobs associated with AI datacenters, such as tensor parallelism (TP), DVFS, and LLM Workload Characteristics.
Section ~\ref{3} describes the cooling and computing characteristics of the AI datacenter.
Section~\ref{4} proposes our hierarchical control framework for both cooling and workload scheduling with a dispatch algorithm.
Section~\ref{5} presents a comprehensive performance analysis using simulation and real-system experiments, which validate the effectiveness of the integrated cooling-computing control.
Finally, Section~\ref{6} concludes the paper and outlines directions for future work.

\section{BACKGROUND}
\label{2}
\subsection{Datacenter Cooling Systems}
Cooling systems are major determinants of the thermal behavior of computing components such as GPUs and CPUs in the datacenter. Due to their importance in thermal management and energy efficiency, there are significant opportunities to enhance datacenter performance through optimized cooling strategies. Therefore, it is important to determine their working characteristics to facilitate co-optimization of computing and cooling power consumption. Datacenters utilize air and liquid cooling to control the massive heat loads generated by high-performance GPUs \cite{reff23,reff24,reff26}. Air cooling remains the most widely used approach due to its relatively low operating and maintenance costs~\cite{reff27}. Heat is removed by circulating cold air through the servers, which can be deployed at the level of the room, row, or rack. Room-level systems pump cold air through the raised floor plenum, but are subject to problems of cold air bypass and return of hot air, resulting in reduced efficiency in the cooling process~\cite{reff28}. 

According to ASHRAE guidelines, proper inlet and outlet air temperature must  be maintained to ensure stable operation of IT equipment and to avoid hardware failure~\cite{reff32}. To improve cooling effectiveness, several key parameters can be actively controlled, including supply air temperature, inlet airflow rate, cold inlet temperature, and return temperature.  These parameters must be dynamically tuned to track the time-varying AI workloads while ensuring thermal safety. Effective control of these parameters is possible with an advanced control scheme. PID (Proportional–Integral–Derivative) control is commonly used in temperature control in a datacenter~\cite{reff33}. Despite its simplicity, PID controller only relies on the previous error values for control, and it is incapable of considering forecasts for the error values. Hence, it can be slow and inaccurate in its responses~\cite{reff34}. In addition,  PID controllers cannot incorporate optimization objectives. While model predictive control (MPC) leverages the knowledge obtained from the system model to predict its possible behaviors in the future~\cite{reff35}.  Modern datacenter cooling systems support fine-grained control through programmable mechanisms such as fan-speed modulation and chilled-water valve adjustment. Recent studies have demonstrated that advanced predictive and learning-based control strategies can be effectively deployed in practical datacenter environments. In \cite{reff35b}, a model-based control framework was applied to datacenter cooling systems and shown to provide effective thermal regulation at sub-minute timescales. Similarly, \cite{reff35c} proposed an LSTM-based predictive control strategy that dynamically adjusts computer room air conditioning (CRAC) operating parameters to improve energy efficiency.  With real-time temperature prediction and optimization of cooling systems, MPC has been extensively used in the field of traditional CPU-based datacenter energy savings and thermal management. In~\cite{reff36}, an objective function of the MPC was formulated that included operating costs, which  reduced energy consumption.  In ~\cite{reff37}, a data-driven subspace predictive control method was designed for air-cooled traditional datacenters. 
An economic MPC was proposed for room-based cooling systems in \cite{reff38,reff39}. However, many of these methods were mainly adapted to CPU-centric datacenters, which have slower workload dynamics and fixed airflow actuation. Such assumptions do not hold for AI/LLM inference, which is highly time-varying and exhibits nonlinear performance–power behavior.

\subsection{LLM Serving and Resource Management}

Existing studies have explored efficient LLM serving from different perspectives. $\mu$-Serve~\cite{reff17b} reduces GPU power consumption through power-aware serving and GPU frequency management techniques, while DynamoLLM~\cite{reff11} improves the efficiency of LLM inference clusters through workload forecasting, request characterization, tensor-parallelism adaptation, dynamic resource allocation, and workload-aware scheduling. By explicitly leveraging LLM workload characteristics, these approaches improve computing-resource utilization and energy efficiency. However, they primarily focus on computing-side optimization and do not explicitly model thermal dynamics or optimize cooling-system operation. TAPAS~\cite{reff18} incorporate thermal and power awareness into workload placement, request routing, and virtual machine reconfiguration decisions. However, TAPAS thermal and power information is used mainly to support scheduling and placement decisions, while cooling-system operation remains unmanaged.

\subsection{Tensor Parallelism and Thermal Constraints} 

As the computational requirements for LLM inference remain high in terms of memory, large models are generally split into multiple GPUs using a concept referred to as model parallelism in LLMs for efficient execution. There exist two variants for model parallelism, namely pipeline parallelism (PP) and tensor parallelism (TP)~\cite{reff40}. In PP, the models' layers are assigned to different GPUs in such a way that each GPU is only responsible for processing a separate stage within a specific layer, with communication between only two successive stages.  In the other variant, called tensor parallelism (TP), computations within each layer are partitioned across multiple GPUs, requiring inter-GPU communication and synchronization at every layer~\cite{reff41}. Because most open-source LLMs can be accommodated on multiple GPUs within a single server~\cite{reff42}, this paper focuses on the widely used TP configurations. Although choosing a smaller TP reduces power consumption, the computational workload for each separate GPU is also higher. Increased usage for each GPU leads to increased heat production by single device, possibly increasing the temperature with subsequent thermal throttling with reduced power~\cite{reff18}. Considering these operational challenges, TP decisions must incorporate temperature constraints to achieve efficient datacenter operation.

\subsection{DVFS and LLM Workload Characteristics}

DVFS is also widely utilized in standard CPU-based datacenters to save power under stable and predictable workloads~\cite{reff43}. DVFS holds the promise of efficiently reducing dynamic power consumption by lowering the operating frequency/voltage of processors during periods of low utilization without compromising performance. In contrast, the workload of LLM serving has highly stochastic characteristics with variations in request sizes influenced by the input/output token lengths. There is also a significant variation in the computational requirements for each request, leading to fluctuations in GPU usage, power usage, and heat generation over a short period of time. As a result, traditional fixed GPU frequency approaches in AI datacenters may not be efficient~\cite{reff11}. To overcome this challenge, DVFS of LLM inference needs to adapt dynamically according to the real-time workload pattern of the arriving requests, in addition to considering the temperature factors.

\subsection{Performance Metrics}

Most prior studies evaluated the latency of LLM inference using the time to first token (TTFT) and the time between tokens (TBT)~\cite{reff44,reff45}. In this work, end-to-end request-level performance is evaluated using latency. Additionally, in terms of measuring the energy consumption for each GPU (in watt-hours, Wh), together with the average GPU temperature during the inference phase, we intend to evaluate the power efficiency and thermal sustainability for different levels of workload, respectively. Cooling performance is evaluated using  cooling energy in conjunction with temperature metrics, including the server return temperature and the inlet (cold-aisle) temperature. In this work, we aim at optimizing the overall AI datacenter performance by integrating both LLM serving and thermal management metrics.

\section{Cooling and Computing Characteristics of AI Datacenters}
\label{3}
In this Section, we present the complete modeling and forecasting framework used in our cooling–computing control system. Our goal is to clearly present the thermal dynamics, power modeling, and IT-load forecasting, and explain why each component is needed and how they connect to the proposed hierarchical control framework. 

\begin{figure}[t]
  \centering
  \includegraphics[width=0.98\linewidth,
    clip, trim=50 50 150 5]{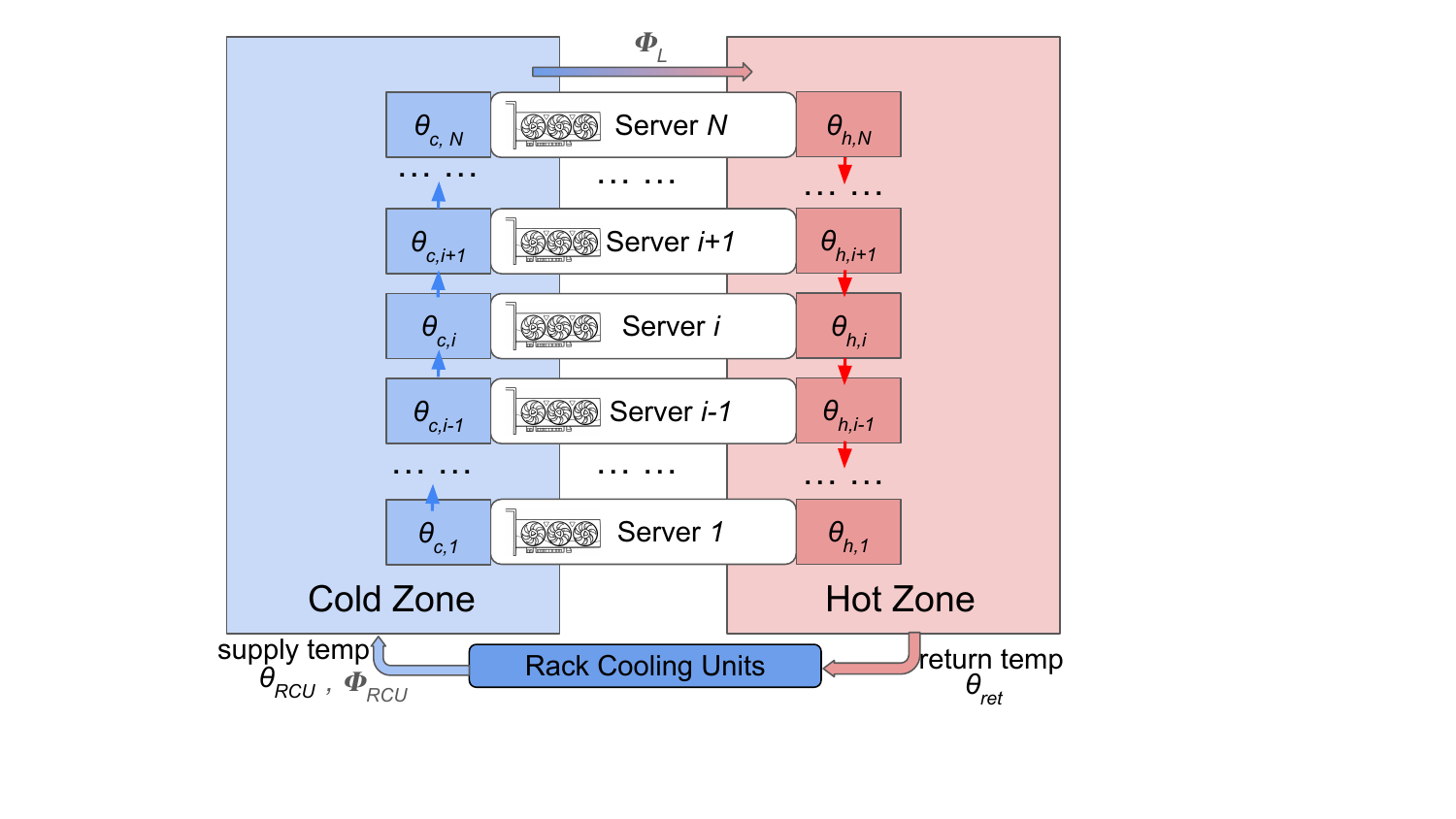}
  \caption{ Schematic of airflow in the  hot and cold zone, with cooling from Rack Cooling Units (RCU).}
  \label{fig:zonal}
\end{figure}
\subsection{General Framework for Rack-Based Air Cooling}
\label{abc}
The rack-level thermal and cooling models used in this work are adapted from the framework presented in \cite{reff48}. These models describe the thermal dynamics of the cold zone, server exhaust air, hot zone, and cooling unit, and are integrated into the proposed hierarchical control framework.
Rack-based cooling is widely adopted in modern datacenters to compensate for limitations associated with   room-/row-based cooling. In this design, rack cooling units (RCUs) are integrated within single server racks, usually at the bottom or at the back. RCUs supply conditioned air to the cold aisle, drawn through server components such as GPUs and memory modules~\cite{reff46}. The heat-exhaled air is released into hot aisle and fed back to the RCU to be cooled once more, with a closed-loop airflow cycle.

To model the thermal dynamics, a zonal modeling method is utilized which divides the rack into distinct thermal zones, generally categorized as the inlet and  outlet zone. The assumption holds that every zone is subject to the same thermophysical properties, namely density and specific heat~\cite{reff47}. The model is governed through mass conservation principles that account for airflow and convective heat transport, with energy balance relations that describe thermal relations between zones.

\subsection{Temperature Field Model and Power Model Development}
\label{def}
The thermal dynamics per-rack for the datacenters is built based on energy conservation laws~\cite{reff48}. Each rack (with server  $i=1,...,N$) is divided into three temperature zones with $\theta_{c,i}$, $\theta_{s,i}$, and $\theta_{h,i}$ denoting the temperatures for the cold, server exhaust, and hot zones, respectively. Let $\Phi_{L}$, $\Phi_{i}^{OH}$, $\Phi_{s,i}$, $\Phi_{\mathrm{RCU}}$ and $\Phi_{i}^{OC}$ denote the leakage flow rate between the hot and cold zones, the recirculated flow between the cold zone  and the hot zone, the server fan flow, the RCU supply flow, and the cold-to-cold coupling flow, respectively.  The RCU supply temperature is denoted by $\theta_{\mathrm{RCU}}$, and $b_i$  indicates the distance from the RCU. Each server contains  $m$ number of  GPU, whose power consumption is modeled by $P_{\mathrm{GPU}}(f(t),u(t))$ as a function of the operating frequency of the GPU $f(t)$ and the utilization of the workload $u(t)$. Given air density $\rho$, heat capacity $c_p$, cold zone volume $V_c$, hot zone volume $V_h$, and server thermal capacitance $C_{th}$, we can model the thermal dynamics for the cold zone, server exhaust and hot zone. The cold-zone dynamics are
\begin{equation}
\begin{split}
\dot{\theta}_{c,i}(t) &= \frac{1}{\rho c_p V_c} \Big( 
\underbrace{b_i \Phi_{RCU} 
\rho c_p \theta_{RCU}}_{\text{RCU supply}}
+ \underbrace{\Phi^{OC}_{i-1}\rho c_p \theta_{c,i-1}}_{\text{upstream inflow}}
 \\
&+ \underbrace{\Phi_L \rho c_p \theta_{h,i}}_{\text{leakage}} - \underbrace{(\Phi^{OC}_i + \Phi_{s,i}) \rho c_p \theta_{c,i}}_{\text{outflow}} \Big),
\label{eq:cold_zone}
\end{split}
\end{equation}

The server exhaust temperature is expressed as
\begin{equation}
\dot{\theta}_{s,i}(t) = \frac{1}{C_{th}} 
\Big( \underbrace{\Phi_{s,i}\rho c_p(\theta_{c,i} - \theta_{s,i})}_{\text{convective exchange}}
+ \underbrace{m P_{GPU}(f(t),u(t))}_{\text{server power}} \Big),
\label{eq:server_temp}
\end{equation}

Finally, the hot-zone dynamics are
\begin{equation}
\begin{split}
\dot{\theta}_{h,i}(t) &= \frac{1}{\rho c_p V_h} \Big( 
\underbrace{\Phi_{s,i}\rho c_p \theta_{s,i}}_{\text{exhaust inflow}}
- \underbrace{\Phi_L \rho c_p \theta_{h,i}}_{\text{leakage to cold}}
+ \underbrace{\Phi^{OH}_{i-1}\rho c_p \theta_{h,i-1}}_{\text{upstream inflow}} \\
&\quad - \underbrace{\Phi^{OH}_i \rho c_p \theta_{h,i}}_{\text{recirculation out}} \Big),
\end{split}
\label{eq:hot_zone}
\end{equation}

\subsubsection*{Thermal Load Extraction}
In addition to the rack thermal dynamics, we model the cooling power consumption of the RCU. Once we collect measurements of the temperature of each hot zone $\theta_{h,i}$, the return temperature $\theta_{\text{ret}}(t)$ is
\begin{equation}
\theta_{\mathrm{ret}}(t) = \frac{1}{N} \sum_{i=1}^{N} \theta_{h,i}(t);
\end{equation}
then the heat removed by the RCU can be calculated as
\begin{equation}
Q_{\text{load}}(t) = \rho \,\Phi_{RCU}(t) c_p \big(\theta_{\text{ret}}(t) - \theta_{\text{RCU}}(t)\big).
\end{equation}

\subsubsection*{Cooling Efficiency Metric}
The efficiency of the RCU is quantified through the coefficient of performance (COP), represented as a quadratic function of the outlet temperature. The cooling-efficiency model is adopted from prior datacenter cooling literature~\cite{reff50}:
\begin{equation}
COP\big(\theta_{\text{RCU}}(t)\big) = \alpha_0 + \alpha_1 \theta_{\text{RCU}}(t) + \alpha_2 \theta_{\text{RCU}}^2(t),
\end{equation}
where $\alpha_0, \alpha_1, \alpha_2$ are the empirical coefficients.  
The power drawn from the cooling source is proportional to the thermal load and inversely proportional to the COP:
\begin{equation}
P_{\text{c,src}}(t) = \frac{Q_{\text{load}}(t)}{COP\big(\theta_{\text{RCU}}(t)\big)}.
\end{equation}
The fan power consumption model is adopted from   datacenter
cooling literature\cite{moazamigoodarzi2020}:  
\begin{equation}
P_{c,\text{fan}}(t) = \delta_0+ \delta_1 \Phi_{RCU}(t) + \delta_2 \Phi_{RCU}^2(t),
\end{equation}
where $\delta_0, \delta_1, \delta_2$ are empirical parameters that reflect the performance of the fan, and $P_{c, \text{fan}}$ is the fan power in W.
The physical validity of the adopted fan-power model over the controller's
operating airflow range is further evaluated in
Appendix \ref{app:fan_model_validation}. The validation confirms that the
predicted fan power remains positive and increases monotonically with
airflow throughout the operating region considered in this work.
 \subsubsection*{Cooling Power Model}

 The cooling power model is adapted from prior datacenter cooling literature~\cite{reff37}. Consequently, the total consumption of cooling power is expressed as the sum of the cooling-source power $P_{c,\text{src}}(t)$ and the fan power $P_{c,\text{fan}}(t)$:
\begin{equation}
P_{\text{cooling}}(t) = P_{c,\text{src}}(t) + P_{c,\text{fan}}(t).
\end{equation}

$P_{c,\text{src}}(t)$  and $P_{c,\text{fan}}(t)$ are main factors defining the power required to remove heat from the GPU rack.
The main source of thermal generation originates from the computing components themselves. In modern AI datacenters, the GPU is the dominant contributor to total IT power,
and its power consumption directly affects the heat released into the surrounding environment. Hence, an accurate model of GPU power consumption as a function of its operating conditions (e.g., working frequency, GPU utilization rate) is essential to characterize the coupling between computing workloads and thermal dynamics. In this work, to characterize GPU power consumption, we adapt the GPU power model from~\cite{reff51}. This model expresses GPU power as a function of operating frequency and utilization verified by real experiments. Our model parameters are identified using telemetry collected from NVIDIA Tesla V100 and A100 GPU during LLM inference experiments:
\begin{equation}
P_{GPU}(t) =   a_{3} (f(t)-f_{ref} )u(t) + a_{2} (f(t) -f_{ref}) + a_{1} u(t) + a_{0},
\end{equation}
where $u(t)$ is the normalized token workload and $f$ is the GPU frequency. The parameters
$a_0,a_1,a_2,a_3$ are coefficients. The GPU frequency is centered around $f_{\mathrm{ref}}=1200$. 

The GPU temperature is adapted from~\cite{reff18}:
\begin{equation}
\dot{\theta}_{GPU_{i}}(t)
=
\beta_0\big(\theta_{c,i}(t)-\theta_{GPU_{i}}(t)\big)
+
\beta_1 P_{GPU}(t)
+
\gamma ,
\label{eq:gpu_temp}
\end{equation}

where $\theta_{GPU_{i}}$ denotes the GPU temperature, $\theta_{c,i}$ is the cooling-air temperature, and $P_{GPU}$ represents the GPU power consumption. The parameters $\beta_0$, $\beta_1$, and $\gamma$ are the model coefficients identified from the experimental data.  $(\theta_C-\theta_{GPU})$ captures the thermal difference between the GPU and its cooling, while the power term models workload-dependent heat generation.

\subsection{Model Parameter Identification and Validation}

 To ensure model accuracy, the parameters of the GPU power model, and  GPU 
temperature model were identified through regression
analysis using telemetry collected from experimental measurements on NVIDIA
GPUs during LLM inference workloads. The resulting identified coefficients
along with their corresponding physical units and parameter descriptions,
are reported in Table~\ref{tab:gpu_model_params}. The GPU frequency is centered around $f_{\mathrm{ref}}=1200$ MHz to define
the regression relative to a representative operating point and improve
the interpretability of the identified coefficients.

\begin{table}[H]
\centering
\caption{Identified GPU power and temperature model parameters. 
The fan power model is adopted from \cite{moazamigoodarzi2020}.}
\label{tab:gpu_model_params}

\small
\setlength{\tabcolsep}{4pt}
\renewcommand{\arraystretch}{1.05}

\resizebox{\linewidth}{!}{%
\begin{tabular}{cccc}
\toprule
\textbf{Parameter} & \textbf{Value} & \textbf{Unit} & \textbf{Description} \\
\midrule

$a_0$ & 138.866 & W 
& Constant GPU power term \\

$a_1$ & 50.556 & W 
& Normalized utilization coefficient \\

$a_2$ & 0.07613 & W/MHz 
& GPU frequency coefficient \\

$a_3$ & 0.13060 & W/MHz 
& Frequency--workload interaction coefficient \\

\midrule

$\beta_0$ & 0.0395 & s$^{-1}$ 
& Thermal relaxation coefficient \\

$\beta_1$ & 0.001579 & $^\circ$C/(W\,s) 
& Power-to-temperature-rate coefficient \\

$\gamma$ & 0.684887 & $^\circ$C/s 
& Residual thermal-rate term \\

\midrule

$\delta_0$ & 480 & W 
& Constant RCU fan-power term \\

$\delta_1$ & -3073 & W/(m$^3$/s) 
& Linear RCU airflow coefficient \\

$\delta_2$ & 6031 & W/(m$^3$/s)$^2$ 
& Quadratic RCU airflow coefficient \\

\bottomrule
\end{tabular}%
}
\end{table}

To assess model accuracy, the identified models were validated using datasets that were not utilized during parameter estimation. Model performance was evaluated using the  root mean square error (RMSE) and mean absolute error (MAE). The validation results are presented in Table \ref{tab:model_validation_summaryy}.  The COP model coefficients adopted from \cite{reff50} are summarized in Table~\ref{tab:cop_params}. The parameter-identification procedure, model-validation methodology and the corresponding validation results including the GPU power response
with respect to frequency and utilization and the steady-state GPU
temperature response with respect to utilization, are  provided in Appendix~\ref{ModelVal}- \ref{app:fan_model_validation}
.
\begin{table}[H]

\centering

\caption{model  validation results.}

\label{tab:model_validation_summaryy}

\begin{tabular}{lcc} 

\toprule

Model & RMSE & MAE  \\ 

\midrule

GPU Power Model & 8.358 W & 6.523 W  \\

GPU Temperature Model & 1.51 $^\circ$C & 1.01 $^\circ$C  \\

\bottomrule

\end{tabular}

\end{table}

\begin{table}[H]
\centering
\caption{COP model coefficients adopted from \cite{reff50}.}
\label{tab:cop_params}
\begin{tabular}{ccc}
\toprule
Parameter & Value & Description \\
\midrule
$\alpha_0$ & 0.458  & Baseline COP coefficient \\
$\alpha_1$ & 0.0008 & Linear temperature coefficient \\
$\alpha_2$ & 0.0068 & Quadratic temperature coefficient \\
\bottomrule
\end{tabular}
\end{table}

\begin{table}[b]
\centering
\caption{Performance metrics with different configurations of TP units for varying levels of total tokens, with 8$\times$Tesla V100 GPUs, on the Llama-2 7B model inference workload. Power denotes the aggregate
power of the GPUs participating in each TP configuration, and temperature
denotes the mean temperature across the participating GPUs.}
\label{tab:tp_metrics}
\renewcommand{\arraystretch}{0.9}   
\setlength{\tabcolsep}{3pt}         
\begin{tabular}{c ccc ccc ccc}
\toprule
\multirow{2}{*}{\textbf{Total Tokens}} &
\multicolumn{3}{c}{\textbf{Latency (s)}} &
\multicolumn{3}{c}{\textbf{Temperature (°C)}} &
\multicolumn{3}{c}{\textbf{Power (W)}} \\
\cmidrule(lr){2-4} \cmidrule(lr){5-7} \cmidrule(lr){8-10}
& TP2 & TP4 & TP8 & TP2 & TP4 & TP8 & TP2 & TP4 & TP8 \\
\midrule
195k & 0.473 & 0.395 & 0.365 & 54.7 & 50.8 & 47.4 & 145 & 299 & 598 \\
179k & 0.435 & 0.384 & 0.355 & 54.1 & 50.0 & 47.0 & 145 & 298 & 597 \\
177k & 0.423 & 0.378 & 0.355 & 53.6 & 49.6 & 46.6 & 145 & 298 & 596 \\
168k & 0.368 & 0.355 & 0.344 & 53.4 & 49.5 & 46.6 & 145 & 297 & 596\\
150k & 0.355 & 0.279 & 0.233 & 53.2 & 49.0 & 46.6 & 140 & 289 & 581\\
\bottomrule
\end{tabular}

\end{table}

\subsection{Workload Forecasting Model}
\label{fgh}
Precise workload prediction is critical to designing effective allocation of computing resources for large-scale inference clusters for LLM. By foreseeing future inference demands, the system identifies a proactive workload distribution of computing resources to adjust cooling, balance thermal regulation, and maintain performance and energy efficiency as a whole. Moreover, datacenters can prevent abrupt changes in thermal and computing conditions ahead of time, thus attaining SLOs compliance with avoided unnecessary power usage. This predictive capability is especially important for serving LLM deployments, where  token generation needs have strong temporal correlation with  LLM serving characteristics, such as prefill-decoding schemes and transformer architectures. To achieve workload forecasting, we collect 30-minute interval data and train a sequence to sequence long-short-term memory (LSTM) network, which fulfills token workload prediction at the cluster level. At time step $t$, we define
the input vector as $L_t = [n_t, g_t]$, where $n_t$ and $g_t$ denote the aggregated context length and  generated tokens within this interval, respectively. Given the  lookback length $H$,  the next-step token demand is predicted as $\hat{L}_{t+1} = f_{\mathrm{LSTM}}\!\left(L_{t-H+1:t}\right)$ where $L_{t-H+1:t}$ denotes the sequence
$\{L_{t-H+1}, \ldots, L_t\}$. The LSTM can preserve memory over time using its hidden states, and can address short-term variations in token demand together with long-term cycles in token demand. 
The architecture and training settings of the LSTM forecasting model are presented in Table \ref{tab:lstm_setup}. 
The LSTM forecasting model achieves a normalized mean squared error (NMSE) of 0.0392, a normalized mean absolute error (NMAE) of 0.0353, and a mean absolute percentage error (MAPE) of approximately 4\% on the test dataset. Detailed model-validation results, including additional error metrics and prediction plots, are presented in Appendix~\ref{ModelVal}.
\begin{table}[H]
\caption{LSTM workload forecasting configuration.}
\label{tab:lstm_setup}
\centering
\footnotesize
\begin{tabular}{p{2.7cm} p{5.0cm}}
\hline
\textbf{Parameter} & \textbf{Configuration} \\
\hline
Dataset &  Azure LLM inference trace~\cite{azurellm2024}\\
Dataset Split & 80\% train, 10\% validation, 10\% test \\
Model & 2-layer LSTM, 128 hidden units \\
Input & 2-hour lookback window \\
Output & Next 30-min token demand \\
Learning Rate & 0.001 \\
Optimizer & Adam \\
Epochs & 8 \\
Metric & MAPE \\
\hline
\end{tabular}
\end{table}
\subsection{Impact of TP and DVFS on LLM Inference Workloads}
\label{sec:DVFS}
The performance of LLM inference is jointly determined by the workload characteristics, the degree of TP, and the underlying GPU frequency settings.  Table~\ref{tab:tp_metrics} shows the profiling results obtained from our real GPU inference experiment using the LLaMA-2 model~\cite{reff42} with 8X Tesla V100 cluster. It illustrates how TP configurations (TP2, TP4, TP8) and token lengths would impact the power and computing profile. For a given total token size, increasing the TP decreases the load per GPU, consequently decreasing the average inference latency and temperature. While larger TP sizes are also responsible for increasing the power usage. This compromise requires determining the best TP configuration given observed workload and datacenter utilization.
\begin{table}[t]
\centering
\caption{Performance metrics with varying GPU frequencies for three representative workload traces, using  V100 GPUs  for Llama-2 7B inference with TP2. Power denotes the aggregate
power of the GPUs participating in each TP configuration, and temperature
denotes the mean temperature across the participating GPUs.}

\label{tab:freq_metrics}
\renewcommand{\arraystretch}{0.9}   
\setlength{\tabcolsep}{2pt}         
\begin{tabular}{c ccc ccc ccc}
\toprule
\multirow{2}{*}{\textbf{Frequency}} &
\multicolumn{3}{c}{\textbf{3047 Tokens}} &
\multicolumn{3}{c}{\textbf{2373 Tokens}} &
\multicolumn{3}{c}{\textbf{935 Tokens}} \\
\cmidrule(lr){2-4} \cmidrule(lr){5-7} \cmidrule(lr){8-10}
& Lat. & Power & Temp & Lat. & Power & Temp & Lat. & Power & Temp \\
\midrule
1000 & 4.169 &  87.71 & 42 & 3.673 & 68.90 & 42 & 3.546 & 67.17 & 40 \\
1200 & 3.811 &  169.41 & 43 & 3.641& 152.78 & 43 & 3.487 & 77.66 & 40\\
1380 & 3.780 & 204.46 & 45 & 3.623& 187.50 & 44 & 3.466 & 163.87 & 43 \\
1450 & 3.732 & 209.13 & 46 & 3.620 & 193.97 & 45 & 3.465 & 164.17 & 43 \\
1500 & 3.712 & 219.99 & 47 & 3.615 & 194.85 & 45 & 3.463 & 156.58 & 43 \\
\bottomrule
\end{tabular}
\end{table}

Second, Table~\ref{tab:freq_metrics} aggregates the effects of DVFS for workloads with varying token lengths. The request streams are categorized into three different sizes of workload based on the total length of the generated and incoming context token. This classification allows for an ordered investigation of how the workload size interacts with the GPU frequency. The table finds that higher frequency settings always reduce latency, but raise power consumption and GPU temperature. In particular, since token length  directly affects each workload's computation demand, the latency, power, and thermal readings are different for the three classes. The per class effect of DVFS is reported in Appendix~\ref{Perclass DVFS}.

These tables show the subset of  experimental setup utilized to investigate the effects of workload variation for various TP and frequency tuning settings. By classifying incoming requests into small, medium and large sets based on the integrated context and generated token lengths, we can determine how the size of the workload influences computational requirements. The dataset, training configuration, input features, and evaluation metrics used for the DistilBERT workload classifier are presented in Table \ref{tab:distilbert_setup}. For each type of workload, we vary the TP configuration and the operating frequency of the GPU, then record the resulting latency, power usage, and temperature levels. This methodology allows for the determination of the sensitivity of LLM inference performance to the change in size of the workload and hardware control parameters. Overall, these results highlight a clear trade-off: a higher TP and higher GPU frequency reduce latency, but they can increase power draw and temperature. These experimental observations motivate our choice of TP and DVFS as the decision variables in the proposed framework.

\begin{table}[t]
\centering
\caption{DistilBERT workload classifier configuration and evaluation setup.}
\label{tab:distilbert_setup}

\renewcommand{\arraystretch}{0.9}
\setlength{\tabcolsep}{3pt}

\begin{tabular}{p{3.7cm} p{4.7cm}}
\toprule
\textbf{Parameter} & \textbf{Configuration} \\
\midrule
Dataset & Azure LLM inference trace~\cite{azurellm2024} \\

Dataset Split &
80\% train, 10\% validation, 10\% test \\

Model &
DistilBERT \\

Training Epochs &
6 \\

Context Tokens &
Used for training and inference \\

Learning Rate &
$2\times10^{-5}$ \\

Evaluation Metrics &
Accuracy, precision\\
\bottomrule
\end{tabular}
\end{table}

Tables~\ref{tab:tp_metrics} and \ref{tab:freq_metrics} report measurements at
two different profiling granularities.
Table~\ref{tab:tp_metrics} reports interval-averaged metrics. Each
tensor-parallel configuration is profiled over a 5-minute window during which
requests from the Azure conversation trace are continuously served using vLLM
batching. Total tokens denotes the sum of context and generated tokens
processed during the corresponding window. The batching 
settings are held fixed across the TP configurations so that the
tensor-parallel degree is the varied configuration parameter.
For power, the measurements from all GPUs participating in a tensor-parallel
configuration are summed, and the resulting aggregate GPU power is averaged
over the 5-minute profiling window. Thus, the TP2, TP4, and TP8 power values
represent the aggregate power of 2, 4, and 8 GPUs, respectively.
Temperature is reported as the mean temperature across the participating GPUs
and over the same profiling window. Latency is the mean  latency for requests served during the window.
Table~\ref{tab:freq_metrics}, in contrast, reports per-job metrics.
Each entry corresponds to a single request of the stated token length executed
in isolation at the specified GPU frequency using TP2; therefore, no request
batching is involved. Power represents the aggregate power of the two
GPUs participating in TP2, averaged over the execution duration of the
request. Temperature is the mean temperature across the two TP2 GPUs over the
same execution period, and latency is the end-to-end execution time of the
individual request.

As shown  in Table~\ref{tab:distilbert_setup}, the DistilBERT workload classifier uses token-length information as the primary input feature for workload classification.

\section{Hierarchical Cooling and Computing Control}
\label{4}
In this Section, we introduce the design of our hierarchical control architecture, which integrates four levels of decision making to jointly manage computing and cooling parameters in LLM inference. At the upper layer, a cluster-level controller determines how many GPU clusters to activate based on workload forecasts obtained from an LSTM prediction model, updated every 30 minutes. In the second layer, the TP configurations are adjusted every 5 minutes to meet workload demands, while incorporating a novel thermal constraint to prevent overheating. In the third layer, the cooling control is updated every 60 seconds to regulate airflow and inlet temperature for thermal stability. Lastly, at the lowest layer, per-job DVFS scaling is applied, whereby GPU operating frequencies are chosen at run-time to meet latency, energy, and thermal constraints. Through this multi-layered design, the system can achieve coordinated control within various timescales, both the long-term workload variability and the fine-grained system changes.

\subsection{Cluster-Level Control}

At the cluster level, 30-minute intervals are utilized to adjust decisions to decide the number of active servers and the GPU budget subjected to lower layers.
The aim is to allocate adequate computational capacity to process the forecasted token load while avoiding both under-provisioning and excessive idle capacity.
The workload prediction is made based on the LSTM model trained on historical data to predict the sum of tokens in the next half hour. To map that demand to physical resources, we use profiled throughput information from a single TP8 (8-GPU) because it provides the maximum parallelism.  This configuration serves as a reference for capacity estimation, ensuring that cluster-level sizing reflects the maximum processing capacity achievable under full GPU utilization. From these measurements, we obtain the effective 30-minute
capacity \(C_8^{(30)}\). The number of TP groups (servers) required to reliably serve the forecast load is calculated as \(N = \lceil\hat{L}_{30} / C_8^{(30)} \rceil \),
where \(\hat{L}_{30}\) denotes the total token demand predicted by the LSTM during the next 30 minute interval and
\(C_8^{(30)}\) represents  the profiled token-processing capacity (tokens per 30 minutes)  of a TP8 server over the same horizon. 
This profile-anchored sizing translates the forecasted workload into a robust GPU budget, 
which is subsequently passed to the window-level mixed-integer linear program (MILP) 
for fine-grained allocation across TP modes.

\subsection{Window-Level TP Selection}

At the start of each control window, denoted as~$w$ (5 minutes), the system
determines the appropriate TP configuration mix~$m \in \mathcal{M}$, where
$\mathcal{M}$ is the set of candidate GPU parallelism settings, e.g.,
$TP_2$, $TP_4$, etc.
The objective of optimization is to reduce overall power consumption while (i) meeting the anticipated token demand~$\hat{L}_{w}$, (ii) adhering to the specified GPU budget $G_{max}$ and (iii) maintaining compliance with thermal regulation.

The forecast workload $\hat{L}_{w}$ (total tokens in window $w$) is given from an LSTM model that was trained on traces from the past. The decision variable $Y_{m,w}$ is the number of pools of TP$_m$ to be selected for window $w$. The choice of sets $\mathbf{Y}_w = \{Y_{2,w}, Y_{4,w}, Y_{8,w}\}$ is formulated as an MILP with the GPU budget and the temperature constraints.
\begin{subequations}
    \begin{align}
\min_{\mathbf{Y}_{m,w}} \quad & 
    \sum_{m} \sum _{t}{P_{GPU_{m,w,t}}\, Y_{m,w}} \label{eq:obj} \\[0.2em]
\text{s.t.} \quad 
& \sum_{m} C_{m,w}\, Y_{m,w} \geq \hat{L}_w \label{eq:coverage} \\[0.2em]
& \sum_{m} m\, Y_{m,w} \leq G_{\max} \label{eq:budget} \\[0.2em]
& \theta_{GPU,w}(\mathbf{Y}_w,\hat{L}_w,\theta_{c}) \leq \theta_{GPU}^{\max}  \label{eq:thermal} 
\end{align}
\end{subequations}

The pool-level MILP in~\eqref{eq:obj}--\eqref{eq:thermal} determines the
number of TP pools $Y_{m,w}$ assigned to each configuration $m$ during
control window~$w$.
 Designed to minimize the  energy consumption of the GPU ~\eqref{eq:obj}.
The constraint~\eqref{eq:coverage} guarantee that the aggregate capacity of all active pools~$C_{m,w}$ is  adequate to support the expected token demand; The constraint~\eqref{eq:budget} applies the constraint budget of the GPUs $G_{\max}$ (total available GPUs), while~\eqref{eq:thermal} will ensure that the resulting temperature $\theta_{GPU,w}$  from the thermal dynamics in~\eqref{eq:gpu_temp} remains below the maximum allowable limit, i.e., $\theta_{\mathrm{GPU}}(t)\le \theta_{\mathrm{GPU,max}}$, under the selected TP configuration $Y_w$ and the forecast workload $\hat{L}_w$.

Here, $C_{m,w}$, $P_{GPU _{m,w}}$ and $\theta_{GPU,w}$ denote the profiled token-processing capacity (tokens per 5 min), the GPU power in the window, and the temperature that correspond to configuration~TP$_m$, respectively.
When the runtime configuration for the pool is reached, expressed as $\mathbf{Y}^\star_w = (Y^\star_{2,w}, Y^\star_{4,w}, Y^\star_{8,w})$, the runtime controller  uses a smart-switch algorithm and maps the TP pools as desired to the currently idle GPUs. When sufficient idle resources exist, it allows the fast reconfiguration of TP without restarting the serving process or repeating expensive re-sharding. 

Moreover, to design a fair and deterministic dispatch of inference request allocations among the designated multi pools, we utilize a  proportional deterministic dispatch (Algorithm~1). This algorithm assigns tasks to the pools with respect to their evaluated capacities, resulting in an alternating schedule among the pools with respect to their relative weights. Thus, two pools with the same type (e.g., two pools of TP$_4$) are load-balanced by design, with the consequence that job allocation is capacity-aware and predictable for each designated window.
\begin{algorithm}[t]
\caption{Capacity-Proportional Deterministic Dispatch}
\label{alg:dispatch} 
\begin{algorithmic}[1]
\State \textbf{Input:} Active pools $\mathcal{R}'=\{p\}$ with capacities $C_p$; 
       window job list $\{(t_j,n_j)\}_{j=1}^{J_w}$ sorted by arrival time. where $n_j$ denotes the token length of job $j$.
\State \textbf{Output:} Mapping $f: \{1,\dots,J_w\}\to \mathcal{R}'$ (job $\to$ pool)
\State Compute shares $s_p \gets C_p / \sum_{q\in\mathcal{R}'} C_q$
\State Balanced-round $n_p \approx s_p J_w$ to integers $n_p$ 
       such that $\sum_p n_p = J_w$
\State Build a deterministic weighted schedule $S$ by interleaving each pool $p$ 
       exactly $n_p$ times 
\For{$j = 1$ to $J_w$}
    \State $f(j) \gets S[j]$ 
\EndFor
\State \Return $f$
\end{algorithmic}
\end{algorithm}

The Capacity-proportional deterministic dispatch algorithm allocates inference jobs across active GPU pools based on their effective capacities. 
Here, $\mathcal{R}'$ indicates the set of active GPU pools considered in the current scheduling window, and $p$ stands for the index of each of these pools in $\mathcal{R}'$. 
$C_p$ is the effective processing capacity of the pool $p$, defined as the number of tokens processed per window. 
The number of inference jobs that arrive in the current scheduling window is expressed as $J_w$, 
$s_p$ is the normalized capacity share of the pool $p$, and ${n}_p$    denotes the number of jobs scheduled for the pool $p$.  
The function $f: \{1, \ldots, J_w\} \rightarrow \mathcal{R}'$ assigns each job $j$ to a specific group $p$ based on the deterministic schedule determined by the algorithm.
\begin{table}[H]
\centering
\caption{Job classification by input/output token thresholds (DistilBERT-based predictor).}
\label{tab:job_classes}
\setlength{\tabcolsep}{2pt} 
\begin{tabular}{l c c c}
\toprule
\textbf{Query Type} & \textbf{Class} & \textbf{Context Token Length} & \textbf{Generated Token Length} \\
\midrule
Short  & S & $[0,256)$     & $[0,100)$ \\
Medium & M & $[256,1024)$  & $[100,350)$ \\
Long   & L & $[1024,8192)$ & $[350,\infty)$ \\
\bottomrule
\end{tabular}
\end{table}
\subsection{Cooling MPC  Formulation}
\label{subsec:cooling_mpc}
Utilizing the computational power  within a physics-based rack model framework,
we regulate the temperature of the RCU supply and the flow rate.
\(
u_k := (\theta_{\mathrm{RCU}}(k),\, \Phi_{\mathrm{RCU}}(k)),
\)
 reduce cooling power while ensuring thermal safety. Let \(x_k \in \mathbb{R}^{N_k}\) denote the overall thermal state for each active server.
For server \(i\) governed by the zonal energy balance equations 
(Eqs.~\ref{eq:cold_zone}–\ref{eq:hot_zone}),
with \(N_k\) the number of active servers in the time interval \(k\).
We find the discretized dynamics for the sampling time interval  \(\Delta t\) by numerically
integrating the ODE  Eqs. \eqref{eq:cold_zone}-\eqref{eq:hot_zone}
\begin{equation}
x_{k+1} \;=\; f\!\big(x_k,\, u_k,\, d_k\big),
\label{eq:cool_dyn}
\end{equation}
where \(d_k \) is the computing workload.
At each time \(t\), the controller solves the following optimization problem with horizon $N_p$: 
\begin{subequations}
    \begin{align}
\min_{\{u_k\}_{k=0}^{N_p-1}} \quad
& \sum_{k=0}^{N_p-1} 
    P_{\mathrm{cooling}}(k) 
\label{eq:mpc_obj} \\
\text{s.t.} \quad
& x_{k+1} = f(x_k,u_k,d_k), \,k=0,\dots,N_p-1,
\label{eq:mpc_dyn} \\
& \theta_{\mathrm{RCU}}^{\min} \le \theta_{\mathrm{RCU}}(k) \le \theta_{\mathrm{RCU}}^{\max}
\label{eq:mpc_bo} \\
& \Phi_{\mathrm{RCU}}^{\min} \le \Phi_{\mathrm{RCU}}(k) \le \Phi_{\mathrm{RCU}}^{\max} \label{eq:mpc_bounds} \\
& \theta_{\mathrm{c}}(k) \le \theta_{\mathrm{c}}^{\max} 
\label{eq:mpct}\\
  &\theta_{\mathrm{ret}}(k) \le \theta_{\mathrm{ret}}^{\max} 
\label{eq:mpc_inlet_return}\\
& \theta_{\mathrm{GPU}}(k) \le \theta_{\mathrm{GPU}}^{\max} \label{eq:mpc_gpu} 
\end{align}
\end{subequations}

In this cooling MPC,  \eqref{eq:mpc_obj} minimizes the total cooling power on the prediction horizon. 
Constraints \eqref{eq:mpct}--\eqref{eq:mpc_gpu} enforce thermal safety by keeping the temperatures   below their respective limits, while constraints \eqref{eq:mpc_bo}--\eqref{eq:mpc_bounds} impose actuator bounds on the cooling setpoints $(\theta_{\mathrm{RCU}},\Phi_{\mathrm{RCU}})$. In practice, such an MPC framework can be solved iteratively with the fitted cooling dynamics~\eqref{eq:cool_dyn}.

\subsection{Per-Job DVFS with Class-Aware Constraints}

DVFS is a well-established mechanism in GPU power management that dynamically adjusts the GPU clock/voltage  to trade inference performance (latency) for lower power consumption \cite{reff53}, where controllers rely on  known execution time and
workload size.
 However, for LLM inference workloads in the AI datacenter, the output length is inherently uncertain and depends on both the input prompt and the model generation process. While the context-token length is
obtained directly from the LLM tokenizer before inference, the generated-token
length is not known until generation is performed. This makes conventional DVFS policies inadequate, as frequency selection cannot be decided solely from static job parameters. To address this challenge, we introduce a DistilBERT-based job classifier that predicts the workload class (\textit{short}, \textit{medium} or \textit{long}) based on the context token length (Table~\ref{tab:job_classes}). We adopt DistilBERT because it provides accurate workload classification with a relatively small model footprint and low inference overhead, making it suitable for real-time scheduling and control decisions in LLM serving~\cite{refDistilBERT}.   This follows the
general idea of proxy-based output-length prediction in~\cite{qiu2024efficient}.
As shown
in Table~\ref{tab:distilbert_setup}, the DistilBERT workload classifier uses
the context-token length as the input feature for workload classification.
Specifically, the context-token length is available before
inference and is therefore used as the classifier input, whereas the
generated-token length is available only after inference and is used solely
for constructing the ground-truth labels during training, validation, and
testing. The ground-truth workload labels are constructed offline using the
context- and generated-token thresholds defined in Table~\ref{tab:job_classes}. When the two
token lengths fall into different classes, the request is assigned to the
higher workload class. Thus, during online
inference, the classifier predicts the workload class using only the
context-token length available prior to request execution.
 
The classifier is trained using a percentile-based request categorization approach following~\cite{reff11}. Specifically, in the test set, the Short, Medium, and Long classes
constitute approximately $3.6\%$, $39.2\%$, and $57.2\%$ of the samples,
respectively. This classification enables the enforcement of class-specific latency and thermal constraints, allowing frequency selection to be formulated as an optimization problem that is token length–aware and ensures compliance with SLOs. To support workloads with significantly larger context lengths, the framework can be extended with an additional ``extremely-long'' request category by retraining the DistilBERT classifier using traces containing larger context lengths. Therefore, the proposed framework is not restricted to the current three request-length classes and can scale to more skewed or long-context LLM workloads.

For each incoming job $j$, we formulate a binary MILP problem that selects a frequency $f$ from the set  $\mathcal{F} = \{1000, 1200, 1380, 1450, 1500\}$\,\text{MHz}. The DVFS optimization problem minimizes GPU power while respecting both the per-class latency and the GPU temperature limits:
\begin{subequations}
\label{eq:DVFS}
    \begin{align}
\min_{x_{f,j}} \quad & \sum_{f \in \mathcal{F}} P_{GPU}(f,n_j)\, x_{f,j} \label{eq:dvfs_obj}\\
\text{s.t.}\quad
& \sum_{f \in \mathcal{F}} x_{f,j} = 1, \label{eq:dvfs_onehot}\\
& \sum_{f \in \mathcal{F}} lat_{c}(n_j,f)\, x_{f,j} \le lat^{\max}_{c(j)}, \label{eq:dvfs_lat}\\
& \sum_{f \in \mathcal{F}} \theta_{\mathrm{GPU}}(f,n_j)\, x_{f,j} \le \theta_{GPU}^{\max}, \label{eq:dvfs_temp}\\
& x_{f,j} \in \{0,1\},\ \forall f \in \mathcal{F}. 
\end{align}
\end{subequations}

Objective \eqref{eq:dvfs_obj} minimizes the power of the GPU by selecting the optimal operating frequency~$f$ for each job~$j$, subject to several constraints.   Constraint~\eqref{eq:dvfs_onehot} enforces a one-hot selection, ensuring that exactly one frequency is chosen per job. 
Constraint~\eqref{eq:dvfs_lat} guarantee that the job latency at the selected frequency
does not exceed the class-specific latency bound $lat^{\max}_{c(j)}$, 
where $c(j)\!\in\!\{\text{S},\text{M},\text{L}\}$ corresponds to short, medium or long job classes. Constraint \eqref{eq:dvfs_temp}  ensures thermal safety by keeping the GPU temperature below 
its upper threshold $\theta_{GPU}^{\max}$.  
Here, $n_j$ denotes the context token length of the job~$j$, 
while $P_{GPU}(f,n_j)$, $lat_{c}(n_j,f)$, and $\theta_{\mathrm{GPU}}(f,n_j)$ represent the GPU power, latency, and temperature profiles associated with frequency~$f$, respectively.

\begin{table}[t]
\centering
\caption{Parameters of the control algorithm.}
\renewcommand{\arraystretch}{1.35} 
\begin{tabular}{lccc}
\toprule
\textbf{Notation} & \textbf{Value} & \textbf{Unit} & \textbf{Description} \\
\midrule
$T_S$ & 30 & sec & Control sampling time \\
$N_p$ & 2 & -- & Prediction horizon of MPC \\
$\theta_{\mathrm{RCU}}^{\min} $ & 20 & $^{\circ}$C & Lower bound of supply-air temp \\

$\theta_{\mathrm{RCU}}^{\max} $ & 24 & $^{\circ}$C & Upper bound of supply-air temp \\
$\Phi_{\mathrm{RCU}}^{\min} $ & 0.28 & m$^{3}$/s & Lower bound of supply airflow rate \\
$\Phi_{\mathrm{RCU}}^{\max} $ & 1.2 & m$^{3}$/s & Upper bound of supply airflow rate \\
$\theta_{\mathrm{ret}}^{\max} $ & 50 & $^{\circ}$C & Upper limit of return-air temperature \\
$\theta_{\mathrm{c}}^{\max} $ & 27 & $^{\circ}$C & Upper limit of cold-aisle inlet temp \\
$\theta_{\mathrm{GPU}}^{\max} $ & 50 & $^{\circ}$C & Maximum  limit of GPU temperature \\
$K_p$ & 4.5 & -- & Proportional gain of PID controller \\
$K_i$ & 0.18 & -- & Integral gain of PID controller \\
$K_d$ & 0.1 & -- & Derivative gain of PID controller \\
\bottomrule
\end{tabular}
\label{tab:control_params}
\end{table}

The ASHRAE TC 9.9 thermal guidelines recommend maintaining inlet-air temperatures below 27$^\circ$C for air-cooled datacenter equipment \cite{ASHRAE2016}, with supply-air temperatures typically maintained below 25$^\circ$C in high-density GPU deployments. Although the NVIDIA Tesla V100 GPU can operate at die temperatures up to approximately 85$^\circ$C before thermal-throttling mechanisms begin reducing performance \cite{V100Thermal2025}, datacenter operators typically maintain GPU temperatures well below this threshold to preserve performance, reliability, and thermal margin. Recent measurements on production AI accelerator platforms report sustained GPU operating temperatures in the range of approximately 50--72$^\circ$C under air-cooled AI workloads \cite{latif2026cooling}. Accordingly, the supply-temperature bounds and thermal constraints used in this work were selected to maintain GPU temperatures below thermal-throttling limits while keeping  inlet temperatures within recommended datacenter operating conditions.

The proposed controller is designed within a specified operating range of temperature and airflow as shown in Table~\ref{tab:control_params}, allowing stable system cooling. For this purpose, the limits $\theta_{\mathrm{RCU}}$ and $\Phi_{\mathrm{RCU}}$ constrain the operation of the cooling unit within the range, while $\theta_{\mathrm{ret}}^{\max} $, $\theta_{\mathrm{c}}^{\max} $, and $\theta_{\mathrm{GPU}}^{\max} $ enforce the thermal safety of the rack and GPU modules. The PID gains $(K_p, K_i, K_d)$ define the dynamic response of the controller (for a detailed mathematical formulation, see Appendix \ref{PID}).

To evaluate the practicality of the proposed framework, we measured the startup and runtime overhead associated with deploying and operating the LLM serving infrastructure. Table~\ref{tab:overhead} summarizes the measured overheads for creating and initializing an 8$\times$V100 GPU instance running LLaMA-2-7B inference.
\begin{table}[H]
\centering
\caption{Measured startup and runtime overheads of creating and initializing a new 8$\times$V100 GPU instance for LLaMA-2-7B inference serving. }
\begin{tabular}{l c}
\toprule
\textbf{Component} & \textbf{Overhead} \\
\midrule
Create a new virtual machine& $\sim$1.5 min\\
Set up the engine configuration & $\sim$16 sec\\
Model Load (LLaMA-2-7B) & $\sim$ 1.5 min \\
Warmup / KV Cache Initialization on GPU & $\sim$30.753s \\
GPU Frequency Reconfiguration & $\sim$50 ms \\
LLM Inference Task Execution & $\sim$2-3 s \\
\bottomrule
\end{tabular}

\label{tab:overhead}
\end{table}

\section{Numerical Experiments}
\label{5}

\subsection{Evaluation Setup}
We evaluate the proposed joint cooling–computing framework through a one-day simulation on GPU clusters, which captures diurnal workload variations derived from the real-world Azure LLM Inference Trace. The hierarchical controller operates on multiple temporal scales: cluster-level planning every $30$ minutes, pool-level TP scheduling every $5$ minutes, cooling control per minute, and DVFS on per-job level.

Before the DVFS stage, we employ a DistilBERT-based job classification model to predict the output length category from the input context tokens. The classifier achieves $91\%$ accuracy on validation set, allowing the controller to anticipate the expected service duration and select appropriate GPU frequencies that balance latency and power efficiency. 
The
corresponding per-class accuracy values for the short, medium, and long classes
are 51.4\%, 94.5\%, and 91\%, respectively. Although the short class exhibits relatively low classification accuracy, the confusion matrix shows that most misclassified short requests are assigned to the medium class. This represents a conservative overestimation of the workload rather than an underestimation. Consequently, these false-positive classifications allocate sufficient or additional computational resources and therefore do not lead to performance degradation or increased latency. The DistilBERT classifier incurs
an average runtime overhead of 7.48~ms per request. To evaluate whether the use of DistilBERT is justified, we further compare it
with a lightweight MLP classifier using exactly the same input information. DistilBERT achieves
91\% test accuracy and outperforms the lightweight MLP baseline while
maintaining a low runtime overhead.

The DistilBERT classifier is not applied directly to the full LLM prompt. Instead, the serving system tokenizer first computes the context-token length, and the classifier operates on this compact workload representation rather than the complete request text. Consequently, the classifier input remains well within DistilBERT's 512-token limit while still providing sufficient information to categorize jobs into short, medium, and long classes for scheduling purposes.
 The MILP optimization problems for the TP and server allocation are solved using PuLP~\cite{reff55}, while the  cooling control is formulated as an MPC problem solved by SciPy SLSQP~\cite{reff56}. The simulation integrates real GPU profiling data and production-level inference traces, including a one-week Azure workload that covers coding and conversation traces. To further validate the practicality of our approach, we conducted one-hour real system experiments on servers equipped with 8 × Tesla V100 GPUs running the Llama-2-7B model~\cite{reff42}. In addition, we performed an ablation study on an NVIDIA A100 GPU platform using the Llama-2-70B model to evaluate the contribution of individual framework components under a larger-scale LLM workload. The control algorithm is implemented on top of vLLM~\cite{reff57}, allowing dynamic TP configuration, workload scheduling, and DVFS runtime adjustment. The proposed framework is compared against three baselines. The first baseline consists of a single pool operating at non-optimized GPU frequency clocks. The second baseline is DynamoLLM~\cite{reff11}, which performs workload-aware server provisioning, request scheduling, and tensor-parallelism management. However, it does not incorporate the proposed joint cooling-compute control framework, GPU DVFS optimization, or MPC-based cooling control. The third
baseline is Adapted TAPAS \cite{reff18}, a thermal- and power-aware scheduling approach
adapted to our 8-GPU experimental platform. It employs multiple predefined
serving pools with different TP configurations and selects among the available
instances based on their serving load and power/thermal conditions. 
 
GPU power consumption and GPU temperature are obtained through direct measurements using NVIDIA monitoring tools. In contrast, cooling power is not directly measured from the physical cooling infrastructure but is estimated using the cooling-power model. For clarity Appendix~\ref{app sources} also provides a table summarizing the quantities used in the experimental evaluation and clearly distinguishing between directly measured quantities and those estimated using the corresponding models.

 \begin{figure*}[t]
    \centering
    \captionsetup{justification=centering,singlelinecheck=false,font=footnotesize}
    \includegraphics[width=0.9\textwidth]{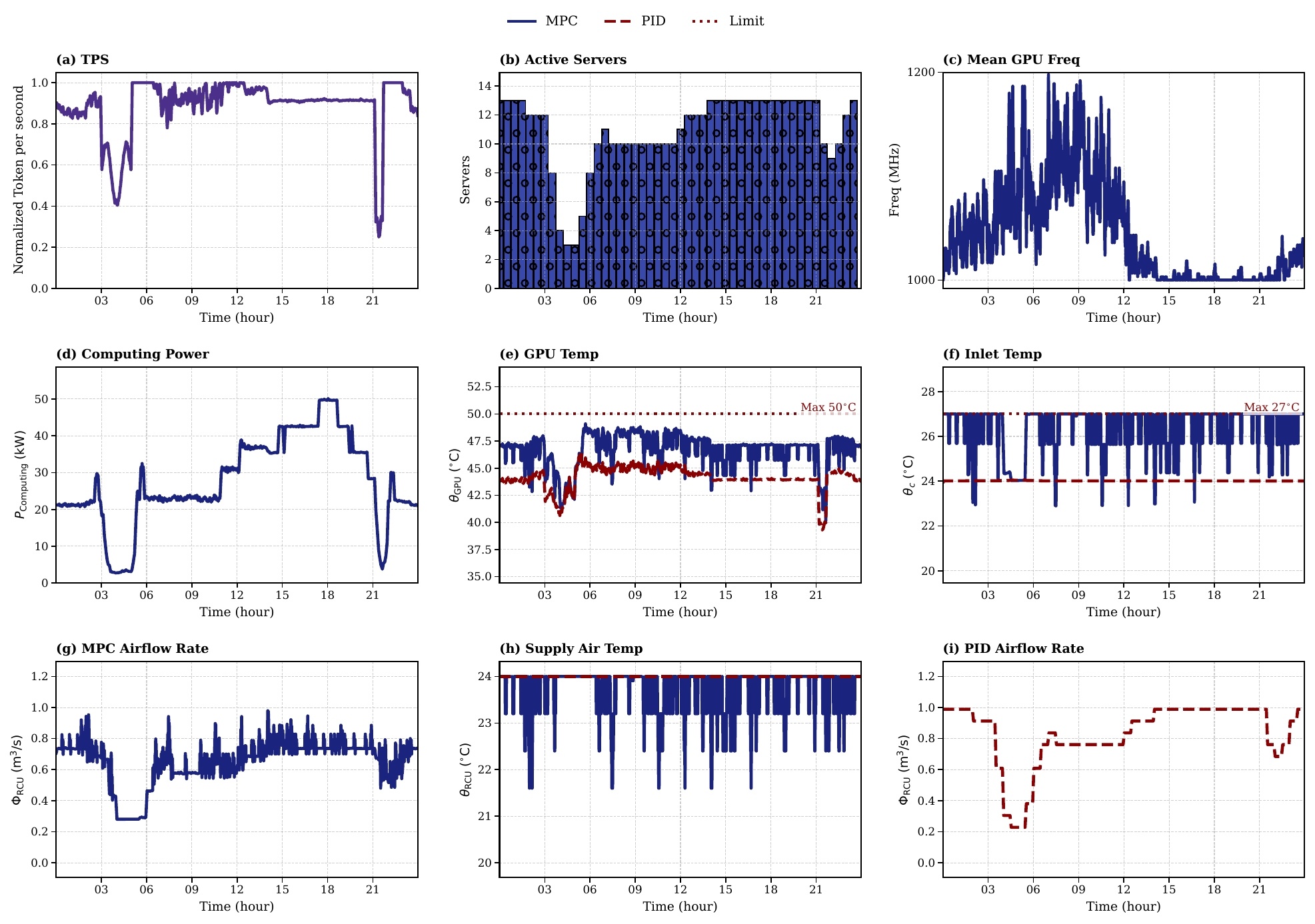}
    \caption{Simulation results of the hierarchical control framework integrating workload utilization, cluster-level provisioning, MILP-based frequency tuning, computing  power consumption, and cooling regulation. The bottom panels show the comparison between PID and MPC controllers in the cooling layer, illustrating the control efforts required by  supply air temperature and airflow rate from both controllers. The reported results are model-estimated.}
    \label{fig:cluster_utilization_provisioning}
\end{figure*}

\begin{figure*}[t!]
    \centering
    \captionsetup{justification=centering,singlelinecheck=false,font=footnotesize}
    \includegraphics[width=0.85\textwidth]{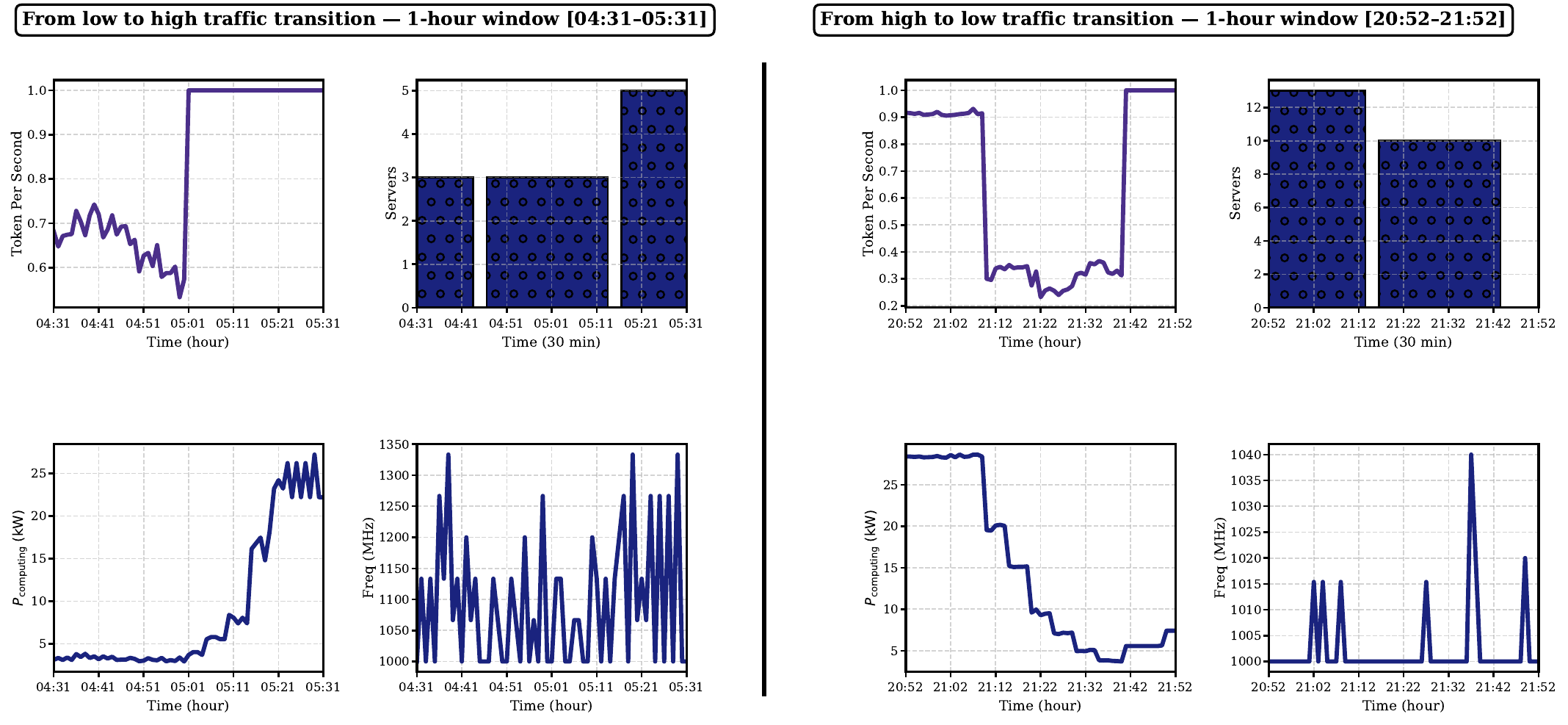}
  \caption{High-resolution one-hour windows illustrating workload transitions from low to high traffic (left panels) and from high to low traffic (right panels). The top row shows workload utilization and the number of provisioned servers, while the bottom row shows computing power consumption and MILP-based GPU frequency selection.}

    \label{fig:cluster_utilization_provisioninglh}
\end{figure*}
\subsection{Hierarchical Control–Based Provisioning}

We simulate the proposed hierarchical control framework using a subset of the utilization profile shown in Fig.~\ref{fig:cluster_utilization_provisioning}\,(a). At the cluster level, the LP-based provisioning controller dynamically  manages the number of active GPU servers according to the workload forecast from the LSTM model.
As illustrated in Fig.~\ref{fig:cluster_utilization_provisioning}\,(b), the controller activates or deactivates GPU servers every 30 minutes to align the computing capacity with the predicted demand.
When off-peak conditions exist, only a few servers are kept online to reduce idle energy consumption, while more servers are brought online as utilization increases.
This adaptive scaling demonstrates that the proposed control  mechanism can effectively balance resource allocation and energy efficiency. 
By maintaining the right number of active GPUs, the proposed controller prevents both under-utilization (too many idle GPUs) and overload (insufficient capacity).  At the DVFS level, an MILP-based frequency-tuning controller determines the
GPU frequency for each workload class to minimize computing power while
satisfying the specified latency and temperature constraints. Fig.~\ref{fig:cluster_utilization_provisioning}\,(c), shows the dynamic scaling of the mean GPU frequency with workload intensity.
This behavior demonstrates the effectiveness of the proposed controller at the   instance-level  in combining compute efficiency with real-time workload dynamics.
The resulting computing power consumption, shown in Fig.~\ref{fig:cluster_utilization_provisioning}\,(d), follows the combined effect of provisioning and frequency control.
Computing power increases with both workload intensity and active server count frequency tuning, and decreases when either utilization or operating frequency is reduced.

\begin{figure*}[b]
    \centering

{    \includegraphics[width=0.9\textwidth]{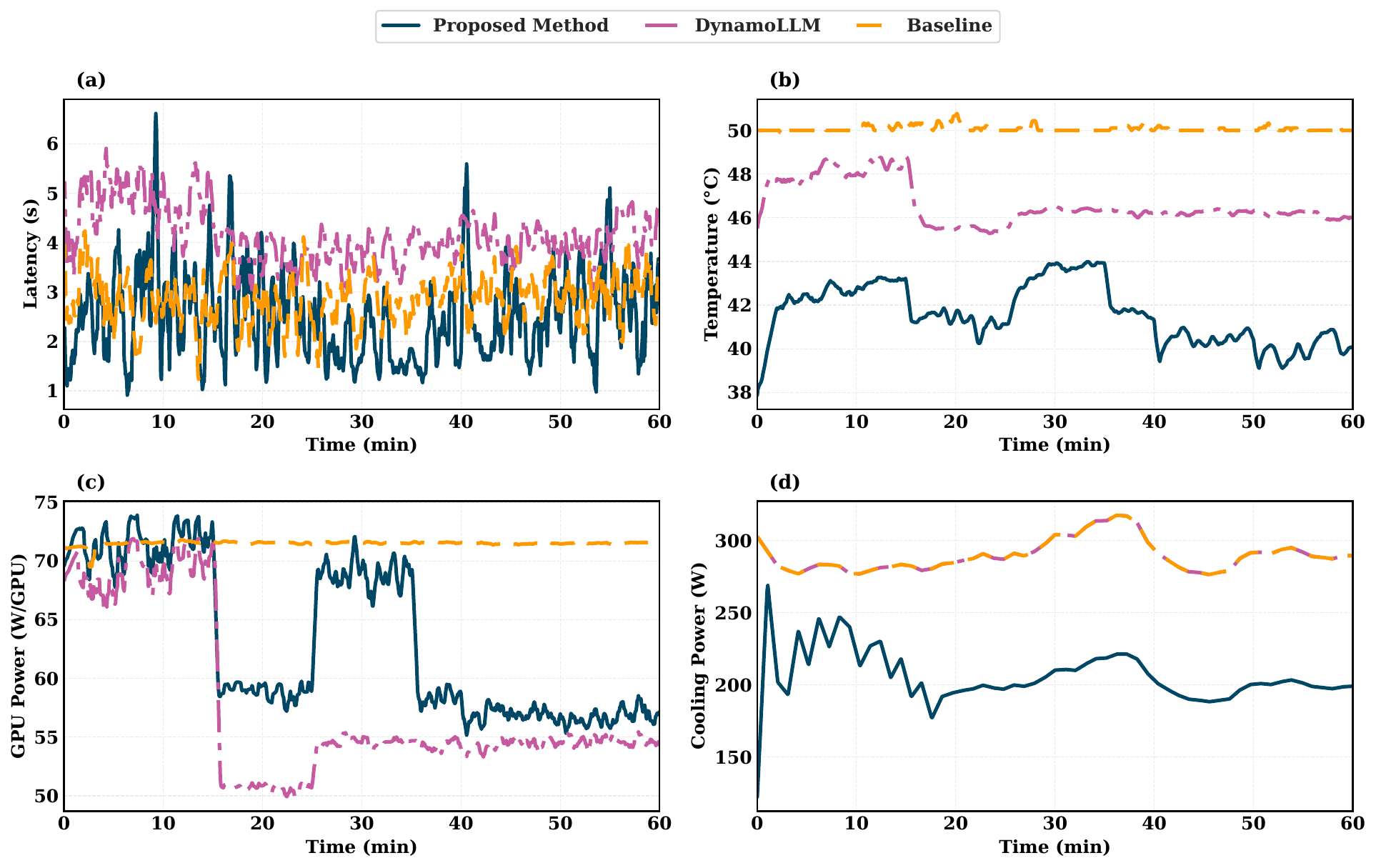}%
        \label{fig:compute_side_eval}
    }

  \caption{Comparison of the proposed hierarchical thermal-aware control framework, DynamoLLM, and the baseline on Llama-2-7B inference deployed on a cluster of eight NVIDIA Tesla V100 GPUs. The plots illustrate (a) end-to-end inference latency, (b) GPU temperature, (c) average per-GPU power consumption, and (d) total cooling power consumption. The proposed framework achieves improved thermal management and lower computing and cooling energy consumption while maintaining comparable inference performance. Inference latency, and GPU computing power are directly measured on the physical GPU
testbed, whereas the cooling power is model-estimated using the corresponding
cooling model.}
     \label{fig:compute_side_eval}
    \vspace{-2mm}
\end{figure*}

\begin{figure*}[t]
    \centering

{    \includegraphics[width=0.98\textwidth]{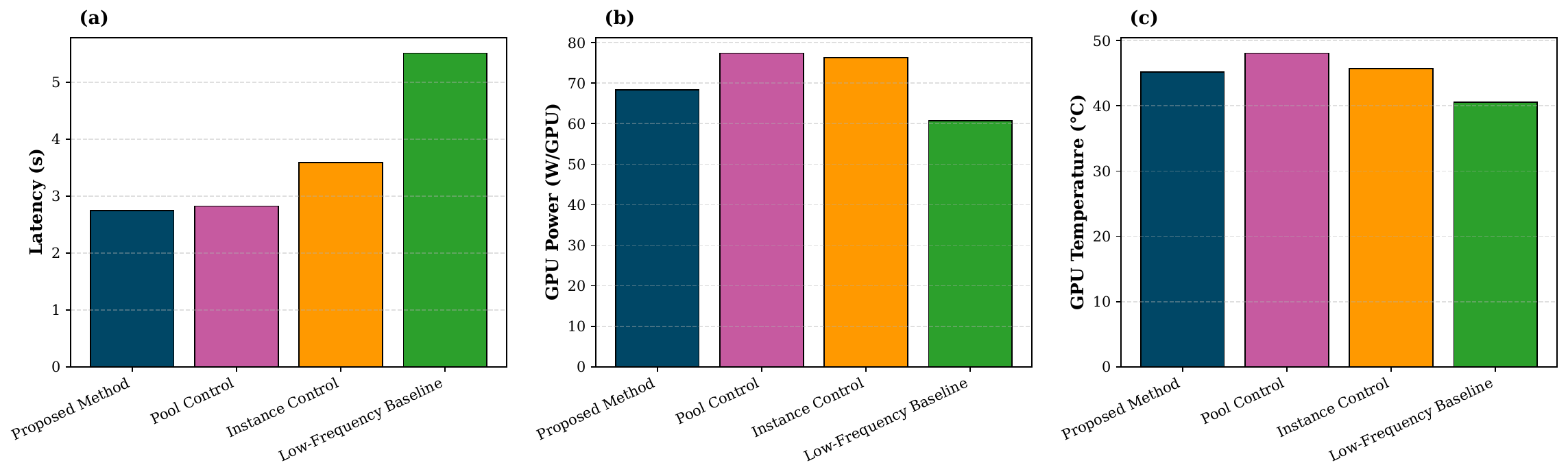}%
    }

  \caption{Ablation study of the proposed hierarchical control framework using LLaMA-2-70B inference workloads on NVIDIA 8 X A100 GPUs. The comparison includes latency, GPU power per GPU, and GPU temperature over the first 20 minutes of the workload. All reported quantities in this figure are
directly measured on the physical GPU testbed.}
     \label{fig:ablation_bar_summary}
    \vspace{-2mm}
\end{figure*}

At the cooling level, an MPC regulates two manipulated variables,  $\Phi_{\mathrm{RCU}}$ and $\theta_{\mathrm{RCU}}$ to control the thermal outputs: temperatures  $\theta_{\mathrm{GPU}}$, $\theta_{\mathrm{ret}}$, $\theta_{c}$, and $\theta_{\mathrm{s}}$.
The objective of MPC controller is to guaranty thermal safety while simultaneously reducing cooling power consumption given GPU TP configuration and incoming workloads.  
 Fig. ~\ref{fig:cluster_utilization_provisioning} (e)-(h) illustrate the performance of the MPC-based cooling controller, which controls in response to the variable intensity of the workload. Fig.~\ref{fig:cluster_utilization_provisioning}\,(e) shows that $\theta_{\mathrm{GPU}}$ stays below the limit of 50\textdegree C throughout the day, which confirms the effectiveness of the proposed controller. As workload and temperature increase, the MPC increases the cooling effort to stabilize $\theta_{\mathrm{GPU}}$. Fig.~\ref{fig:cluster_utilization_provisioning} (f) shows that $\theta_{c}$ increases as the workload is reduced, indicating that $\theta_{c}$ becomes hotter as the workload decreases to minimize the cooling effort and vice versa. This adaptive response from the proposed control system shows that the MPC is working to adjust airflow in the cold aisle according to  thermal and workload variations.  As illustrated in Fig.~\ref{fig:cluster_utilization_provisioning} (g), in the proposed system, $\Phi_{\mathrm{RCU}}$ increases during periods of high utilization, and it is reduced during periods of low utilization to reduce the fan power.  As shown in Fig.~\ref{fig:cluster_utilization_provisioning} (h), the air temperature supplied by the RCU is reduced during periods of high utilization to provide a higher cooling rate and increased during periods of low utilization to reduce the power consumption of the chiller.In addition to the temporal response, the spatial temperature distribution across the rack indicates that servers positioned farther from the cooling unit have slightly higher temperatures due to reduced airflow uniformity and uneven workload distribution. However, every  $\theta_{s}$ is within the defined safety threshold, which confirms effective thermal management under spatially varying conditions (see Appendix~\ref{appendix}). 
To further evaluate the performance, the cooling controller was compared with that of a proposed MPC and a PID controller. In both controls, the workload was the same. The results indicate that the MPC-based cooling strategy significantly outperforms the PID controller in terms of energy savings with around 28\% due to its performance in adapting both $\Phi_{\mathrm{RCU}}$ and $\theta_{\mathrm{RCU}}$ in response to workload variations in advance (see the Appendix~\ref{appendix}). In both cases, the control objective, therefore, is to keep the return-air temperature  below  its set point while minimizing the cooling power.

Fig.~\ref{fig:cluster_utilization_provisioninglh} shows representative,
high-resolution one-hour windows around two operating regimes: a low-to-high
traffic transition (left panels) and a high-to-low traffic transition
(right panels).  Under high traffic times, utilization characteristics show dramatic changes, the LP scheduler activates more servers, and MILP will update GPU frequency to ensure throughput continues. The related MPC thermal control proactively adjusts to the response by increasing $\Phi_{\mathrm{RCU}}$ and reducing $\theta_{\mathrm{RCU}}$ to minimize heat accumulation, thus keeping $\theta_{\mathrm{GPU}}$ below the safety threshold.  Meanwhile, workload and GPU frequencies are low in low-load regimes. MPC also reduces the energy consumption of the cooling by reducing $\Phi_{\mathrm{RCU}}$ and increasing $\theta_{\mathrm{RCU}}$, decreasing the power of the fan and the chiller, while keeping the limits of thermal safety.  On the whole, the side-by-side comparison highlights the advantages of the  proposed coordinated operation of the hierarchical framework. The computation layer scales resources or frequency dynamically with demand.  The cooling layer coordinately adjusts airflow or supply temperature all the time, while achieving joint power efficiency and thermal safety. It is shown that this proposed controller can achieve high performance during peaks and substantial energy savings during troughs without violating thermal or computing constraints.

\subsection{Experimental Evaluation}

We evaluate the effectiveness of the proposed hierarchical control framework using the setup environment described previously. The experiments use three key metrics: GPU power consumption, thermal behavior, and inference latency to assess the performance of the proposed control approach and compare it with  DynamoLLM, Adapted TAPAS  and the baseline setup. These metrics reflect the efficiency of GPU energy usage, operating temperature safety, and real-time inference performance. Figs.~\ref{fig:compute_side_eval}(a)--\ref{fig:compute_side_eval}(d) depict the effectiveness of the proposed framework across the computing and cooling layers under experimental inference workloads.

For a fair comparison, both the proposed framework and DynamoLLM were evaluated using the same Llama-2-7B inference workload deployed on a cluster of eight NVIDIA Tesla V100 GPUs. DynamoLLM dynamically adjusts resource-allocation decisions by selecting tensor-parallel configurations according to the predicted workload demand and available GPU resources. Specifically, DynamoLLM allocates GPUs to different tensor-parallel groups to satisfy workload requirements while improving computing-resource utilization. However, DynamoLLM focuses solely on computing-side optimization and does not explicitly incorporate GPU thermal dynamics, cooling-system models, or cooling-control decisions. In contrast, the proposed framework jointly coordinates workload forecasting, tensor-parallel allocation, GPU frequency management, and MPC-based cooling control. This comparison therefore highlights the benefits of integrating both computing and cooling management within a unified hierarchical control framework.

As shown in Fig.~\ref{fig:compute_side_eval}(a), the total end-to-end latency of the proposed framework remains comparable to  the baseline throughout the experiment, confirming that SLOs are maintained without performance degradation. Fig.~\ref{fig:compute_side_eval}(b) shows the GPU temperature profiles. The proposed framework consistently maintains lower GPU temperatures than both DynamoLLM and the baseline due to its coordinated thermal-management strategy. The average temperature per GPU remains below 44\textdegree C, demonstrating the framework's ability to maintain thermal reliability while reducing energy consumption. Fig.~\ref{fig:compute_side_eval}(c) compares the per GPU power consumption. The proposed hierarchical controller dynamically adjusts tensor-parallel configurations and GPU frequencies according to workload demand and thermal conditions. As a result, it achieves lower power consumption than the baseline while preserving latency performance. During the one-hour experiment, the average reduction in computing energy reaches approximately 24.2\%. To highlight the contribution of the MPC cooling layer, Fig.~\ref{fig:compute_side_eval} (d) shows it modulates $\Phi_{\mathrm{RCU}}$ and  $\theta_{\mathrm{RCU}}$ in coordination with the optimization of the compute side. 

At the pool level, the MILP scheduler dynamically adjusts TP every five minutes based on workload demand and GPU availability (see Table~\ref{tab:tp_config_comparison}). The optimizer switches from TP8 down through TP4 and finally TP2 as workload intensity decreases, reducing energy consumption while maintaining good latency. 
Compared with DynamoLLM, which primarily optimizes computing-resource utilization and inference performance, the proposed framework additionally incorporate GPU power and thermal information into the TP selection process. Consequently, it can select less resource-intensive TP configurations during low-demand periods, reducing GPU power consumption and heat generation. The resulting reduction in thermal load also lowers cooling demand, enabling additional energy savings while maintaining comparable inference performance and satisfying thermal constraints.

\begin{table}[t]
\centering
\caption{TP configuration across five-minute windows W1-W12 for the proposed method and DynamoLLM (LLaMA-2-7B).}
\label{tab:tp_config_comparison}
\resizebox{\columnwidth}{!}{%
\begin{tabular}{llcccccccccccc}
\toprule
\textbf{Method} &
\textbf{TP Mode} &
\textbf{W1} &
\textbf{W2} &
\textbf{W3} &
\textbf{W4} &
\textbf{W5} &
\textbf{W6} &
\textbf{W7} &
\textbf{W8} &
\textbf{W9} &
\textbf{W10} &
\textbf{W11} &
\textbf{W12} \\
\midrule

\multirow{3}{*}{Proposed}
& TP2 & 0 & 0 & 0 & 0 & 0 & 0 & 0 & 0 & 0 & 0 & 2 & 2 \\
& TP4 & 0 & 0 & 0 & 1 & 1 & 2 & 2 & 2 & 1 & 1 & 0 & 0 \\
& TP8 & 1 & 1 & 1 & 0 & 0 & 0 & 0 & 0 & 0 & 0 & 0 & 0 \\
\midrule

\multirow{3}{*}{DynamoLLM}
& TP2 & 0 & 0 & 0 & 1 & 1 & 2 & 2 & 2 & 2 & 2 & 2 & 2 \\
& TP4 & 0 & 0 & 0 & 0 & 0 & 0 & 0 & 0 & 0 & 0 & 0 & 0 \\
& TP8 & 1 & 1 & 1 & 0 & 0 & 0 & 0 & 0 & 0 & 0 & 0 & 0 \\
\bottomrule
\end{tabular}}
\end{table}
\begin{figure*}[b]
    \centering
    \includegraphics[width=0.98\linewidth]{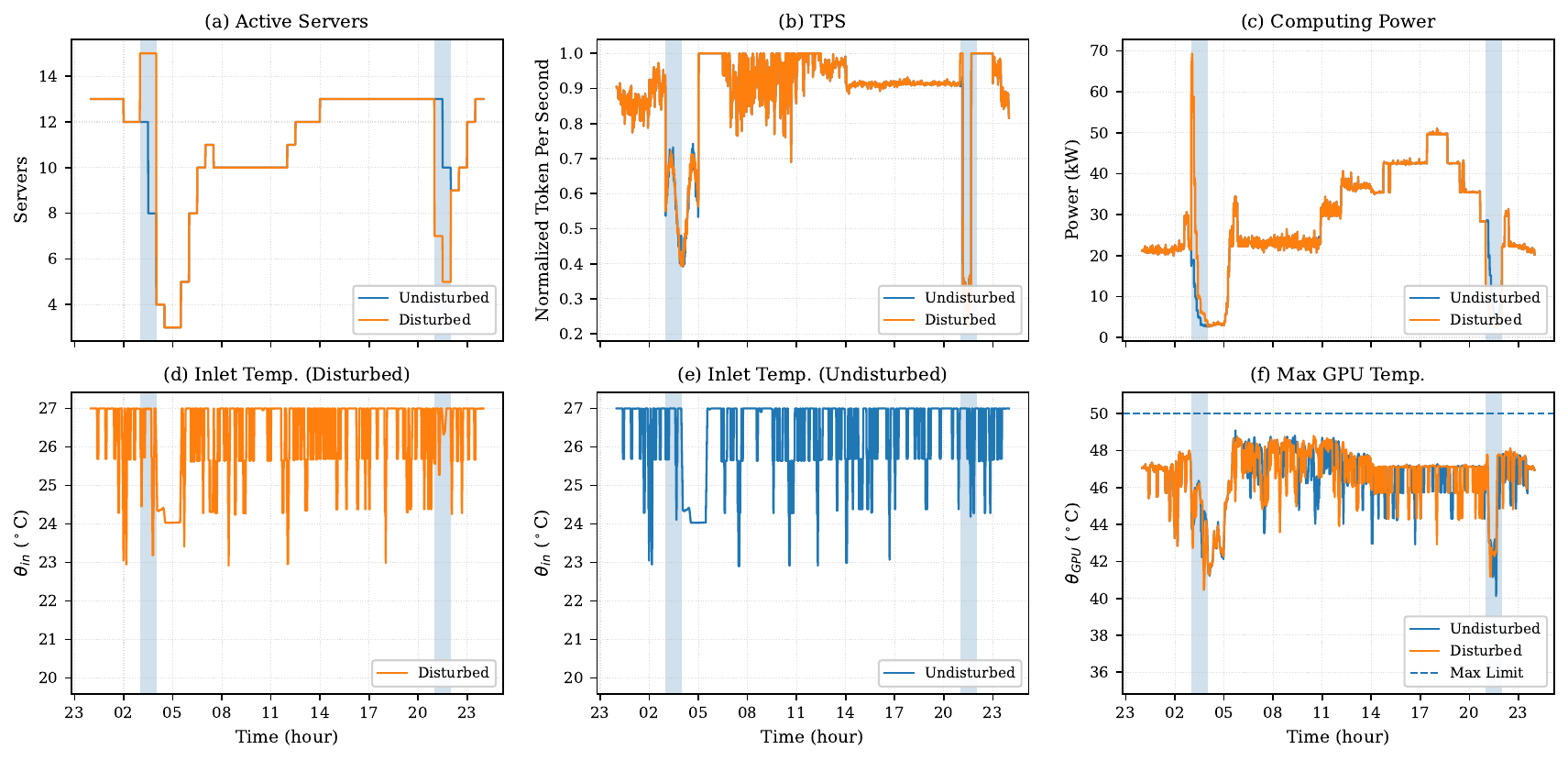}
    \caption{Sensitivity analysis of the proposed framework to workload forecast uncertainty.  Forecast demand was overestimated by  2$\times$  during a low demand period and underestimated by 0.5$\times$ during a peak demand period  for one hour each. All reported quantities in this figure are model-estimated using the
corresponding simulation models.}
    \label{fig:forecast_robustness}
\end{figure*}
Table~\ref{tab:performance_comparisonn} shows that the proposed controller has a reduction in energy in both the cooling and the computing subsystems. The result suggests that  the proposed hierarchical  controller  can efficiently reduce total energy costs while maintaining reliable inference performance.  Table \ref{tab:comparison} compares the proposed framework with DynamoLLM. The proposed method reduces cooling energy consumption by 31.2\%, average GPU temperature by 12.6\%, and latency by 38.4\% compared with DynamoLLM. Compared with DynamoLLM, the proposed method uses 7.6\% more computing energy while reducing cooling energy, average GPU temperature, and latency. When both computing and cooling energy are considered, the total energy
consumption decreases from 91.79~Wh/GPU for the baseline to
67.00~Wh/GPU for the proposed controlled system, corresponding to a
27.01\% reduction in total energy consumption per GPU. Overall, the results demonstrate that jointly coordinating computing and cooling resources can improve thermal efficiency and inference performance compared with a compute-only management approach.

Fig.~\ref{fig:ablation_bar_summary} presents the ablation study of the proposed hierarchical control framework using LLaMA-2-70B inference workloads executed on NVIDIA A100 GPUs. The full proposed method achieves the lowest latency among all methods while maintaining moderate GPU power consumption and GPU temperature. This latency improvement results from the coordinated operation of the pool-level and instance-level controllers. At the pool level, the scheduler dynamically adjusts tensor-parallel configurations according to workload demand and GPU availability, improving resource allocation. At the instance level, workload-aware GPU frequency selection ensures that requests receive sufficient computing resources to satisfy latency constraints while avoiding unnecessary power consumption. The interaction between these two control layers enables the framework to achieve lower latency while maintaining energy efficiency and thermal stability.
\\
Removing the instance-level controller increases both GPU power consumption and temperature, demonstrating the importance of fine-grained frequency adaptation for energy and thermal efficiency. Removing the pool-level controller increases latency, indicating that workload-aware tensor-parallel scheduling contributes significantly to inference performance. The fixed low-frequency baseline achieves the lowest GPU power consumption and temperature, but at the cost of substantially higher latency.

\begin{table}[t]
\centering
\caption{Performance Comparison Between Baseline and Controlled Systems.}
\renewcommand{\arraystretch}{1.15}
\begin{tabular}{lccc}
\toprule
\textbf{Metric} & \textbf{Baseline} & \textbf{Controlled} & \textbf{Improvement} \\
\midrule
Computing energy (Wh/GPU) & 54.8 & 41.6 & \textbf{24.2\%} \\
Cooling energy (Wh/GPU)   & 36.99 & 25.40 & \textbf{31.2\%} \\
\textbf{Total energy (Wh/GPU)} & \textbf{91.79} & \textbf{67.00} & \textbf{27.01\%} \\
Temperature ($^\circ$C)   & 50.1 & 41.6 & \textbf{17.0\%} \\
Latency (s)               & 2.31 & 2.28 & $\approx 0$ \\
\bottomrule
\end{tabular}
\label{tab:performance_comparisonn}
\end{table}

\begin{table}[t]
\centering
\caption{Comparison of the proposed method with DynamoLLM.}
\begin{tabular}{lccc}
\toprule
Metric & DynamoLLM & Proposed & Improvement \\
\midrule
Computing Energy (Wh/GPU) & 38.7 & 41.6 & -7.6\% \\
Cooling Energy (Wh/GPU) & 36.99 & 25.40 & 31.2\% \\
Average Temperature ($^\circ$C) & 47.6 & 41.6 & 12.6\% \\
Latency (s) & 3.70 & 2.28 & 38.4\% \\
\bottomrule
\end{tabular}
\label{tab:comparison}
\end{table}

\begin{figure}[b]
    \centering
    \includegraphics[width=0.99\linewidth]{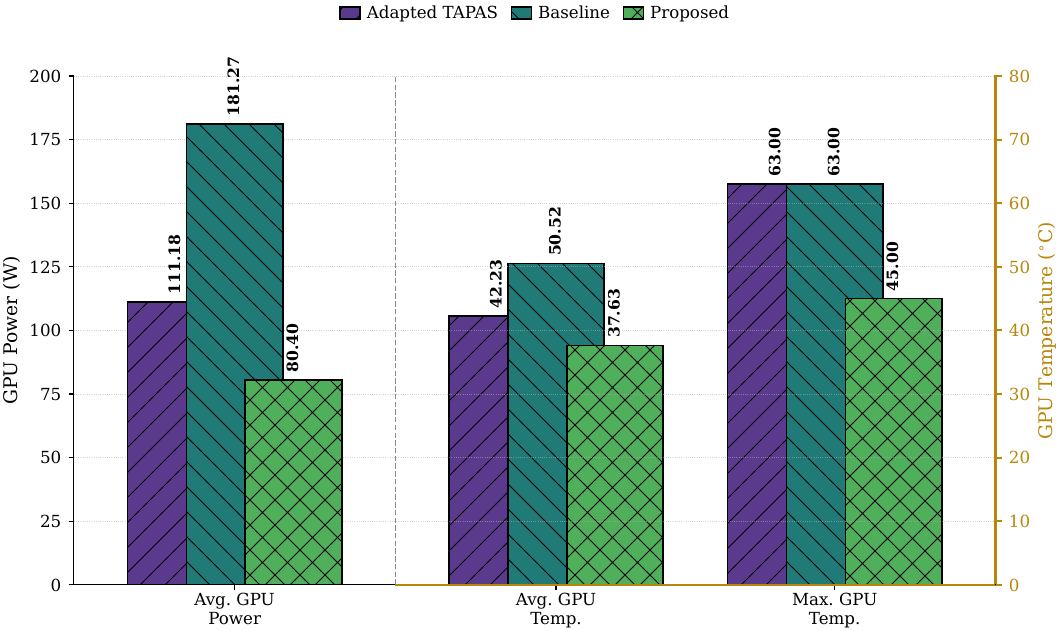}
    \caption{ Power, and thermal comparison of Adapted TAPAS,
    the Baseline, and the proposed approach for Llama-2-7B inference under
    bursty workloads on $8\times$ NVIDIA A100 GPU platform.
    The comparison reports average GPU power,
    average GPU temperature, and maximum GPU temperature.}
    \label{fig:tapas_comparison}
\end{figure}

\subsection{Sensitivity Analysis}

The impact of workload misprediction depends on whether the requested workload
is overestimated or underestimated. In the case of overestimation, the
controller provisions additional computing capacity, ensuring sufficient
resources to serve the arriving workload and potentially reducing queueing and
latency. However, this conservative provisioning may result in energy
inefficiency because more computing resources are activated than necessary.
In contrast, underestimation may lead to insufficient resource allocation,
increased queueing, and degraded latency performance. To mitigate these effects,
the framework incorporates runtime corrective mechanisms. First, the
instance-level controller increases the GPU frequency toward the maximum
allowable value when the observed workload exceeds the prediction. If the
overload persists, the pool-level controller can redirect a subset of requests
to neighboring pools with available resources. Furthermore, requests that wait
beyond a predefined threshold may be temporarily rejected or postponed,
allowing them to be retried once the system load stabilizes.

For the cooling layer, the framework incorporates hard safety constraints on
GPU temperature and cooling control variables. Therefore, despite workload
misprediction, the cooling controller adjusts its operating conditions while
enforcing the prescribed thermal and actuator limits.

To further evaluate the robustness of the proposed framework, we extend the
sensitivity analysis to consider errors in both workload forecasting and
request classification. A 9\% request misclassifications rate is
injected throughout the evaluation horizon. In addition, the forecast workload
supplied to the cluster- and pool-level controllers is multiplied by
\textbf{$2.0\times$} during a low-demand period (03:00--04:00) to emulate
workload overestimation and by \textbf{$0.5\times$} during a peak-demand
period (21:00--22:00) to emulate workload underestimation. Outside these
intervals, the original forecast is used. The actual arriving workload remains
unchanged; only the forecasts and request classifications supplied to the
hierarchical controller are perturbed. Each disturbed interval is compared
with the corresponding  interval of the undisturbed case to account
for temporal variations in workload demand.

Results show that the controller automatically responds to these prediction
errors by adjusting resource allocation, GPU operating frequency, and cooling
conditions.To isolate the energy impact of the injected forecast disturbances, the energy comparison is performed only over the two disturbed one-hour intervals, corresponding to a total disturbance duration of two hours. Considering these two disturbance windows, the disturbed scenario results in an approximately 5\% increase in total energy consumption relative to the corresponding undisturbed intervals.  This change reflects the combined effects of workload
overestimation and underestimation. During overestimation, the controller
activates additional computing resources, thereby increasing computing energy
consumption. In contrast, during underestimation, fewer servers are initially
provisioned, which can reduce computing energy consumption but may adversely
affect request latency.

The prediction errors also affect latency, particularly tail latency. During
the $2.0\times$ overestimation interval, the additional provisioned
capacity reduces the P99 end-to-end latency by approximately
14.7\% relative to the corresponding undisturbed interval. In
contrast, during the $0.5\times$ underestimation interval, fewer
resources are initially provisioned. To capture the potential
carry-over effect of this temporary resource shortage, the reported P99
latency for the underestimation case is evaluated over both the one-hour
disturbance interval and the subsequent one-hour recovery period. 
Over this evaluation window, the P99 end-to-end latency increases by
approximately 20\% relative to the corresponding undisturbed case.

From the thermal perspective,  the injected forecast and classification disturbances have mild effects on GPU temperature. The mean GPU temperature increases by only
0.09\%, while the 95th-percentile temperature remains nearly unchanged.
Similarly, the peak GPU temperature increases only slightly, from
$47.92\,^{\circ}\mathrm{C}$ in the undisturbed case to
$48.12\,^{\circ}\mathrm{C}$ in the disturbed case, corresponding to an
increase of approximately 0.40\%. As shown in Fig.~\ref{fig:forecast_robustness}, the maximum GPU
temperature remains below the prescribed $50\,^{\circ}\mathrm{C}$ thermal
limit throughout the evaluation horizon in both cases. These small thermal
variations demonstrate that the cooling controller effectively compensates
for the effects of workload forecast and classification errors, maintaining
GPU temperatures within the specified thermal constraint despite the
injected disturbances.

\subsection{Sensitivity to Bursty Workloads and Comparison with a
Thermal-Aware Baseline}

To further evaluate the sensitivity of the proposed framework under different
workload and hardware conditions, we extend the evaluation using
bursty Llama-2-7B inference workloads on 
$8\times$ NVIDIA A100 GPU platform. This setting introduces rapid variations
in inference demand and provides an additional multi-GPU configuration for
evaluating the latency, power, and thermal behavior of the proposed framework.

We additionally implement Adapted TAPAS as a representative
thermal- and power-aware scheduling baseline. TAPAS is primarily designed to
enable safe thermal and power oversubscription in GPU-based LLM  \cite{reff18}.
Rather than directly minimizing computing or cooling energy, TAPAS exploits
the available power and thermal headroom while maintaining power, temperature,
cooling-airflow, and workload-performance constraints.

We adapt TAPAS
central scheduling concepts to the 8-GPU environment. Specifically, the GPUs
are organized into three independent vLLM serving instances consisting of two
TP2 instances and one TP4 instance, providing multiple candidate serving
configurations within the available hardware. Adapted TAPAS applies the two
mechanisms that can be directly exercised in this environment:
request routing and instance configuration. 
A feasible serving instance is selected based on its current serving load and available
power/thermal headroom, after which the highest feasible GPU frequency
satisfying the prescribed constraints is applied.

As shown in Fig.~\ref{fig:tapas_comparison}, the proposed approach
achieves the lowest average GPU power of 80.40\,W, compared with
111.18\,W for Adapted TAPAS and 181.27\,W for the
Baseline. This corresponds to approximately 27.7\% lower average
GPU power than Adapted TAPAS.

The proposed framework also provides the lowest average and maximum GPU
temperatures. The average GPU temperatures of Adapted TAPAS, the Baseline,
and the proposed approach are
$42.23\,^{\circ}\mathrm{C}$,
$50.52\,^{\circ}\mathrm{C}$, and
$37.63\,^{\circ}\mathrm{C}$, respectively. Their corresponding
maximum GPU temperatures are
$63\,^{\circ}\mathrm{C}$,
$63\,^{\circ}\mathrm{C}$, and
$45\,^{\circ}\mathrm{C}$, respectively. Therefore, compared with
Adapted TAPAS, the proposed framework reduces the average GPU temperature by
approximately 10.9\% and the maximum GPU temperature by
approximately 28.6\%. The lower maximum temperature is particularly
important under bursty workloads because it demonstrates that the coordinated
controller limits transient thermal buildup during periods of rapidly changing
inference demand.

The proposed framework explicitly coordinates
computing-side resource management with tailored cooling-side control. Consequently,
it achieves substantially lower GPU power and both lower average and peak GPU
temperatures, at the cost of a moderate increase in average inference latency. TAPAS achieves approximately 10\% lower latency than the proposed method. This difference
should not be attributed solely to the performance--energy trade-off introduced
by the proposed control strategy. As discussed above, the request-routing
mechanism of the original TAPAS framework could not be fully reproduced on our
experimental platform. Therefore, in Adapted TAPAS, multiple serving pools are
predefined and requests are assigned among these pools to approximate the
request-routing functionality of the original framework. This implementation
difference can provide readily available serving capacity and thereby reduce
queueing delay relative to the dynamically coordinated resource-management
mechanism used by the proposed framework. Consequently, the observed latency
difference reflects both the different control strategies and the practical
differences in how request routing and resource allocation are implemented. 
Overall, the results demonstrate that the proposed framework maintains a
favorable performance--power--thermal trade-off under the additional bursty
workload and multi-GPU hardware setting.

\section{Conclusion}
\label{6}
This paper tackles the challenge of energy management of AI datacenters, and proposes a hierarchical design for joint computing and cooling control. The proposed framework integrates LP-driven resource allocation, MILP-based LLM scheduling, and MPC cooling control, which is further supported by  an LSTM-based  workload predictor and a latency-aware classifier. Extensive simulations and real system experiments demonstrate up to 24.2\% reduction in per GPU energy consumption, 31.2\% cooling energy savings, and 17\% decrease in mean GPU temperature, resulting in consistent total energy reduction without latency degradation.
Future research will extend the scope further to include advanced liquid and hybrid cooling schemes with the objective of enhancing energy efficiency and temperature stability.

 \bibliographystyle{IEEEtran}
 \bibliography{bib}

\appendix
\input{appendix}


\end{document}

%% file: appendix.tex
\section{}
\label{appendix}

\subsection{Spatial Distribution of Server Exhaust Temperature}
\begin{figure}[H]
    \centering
    
    \includegraphics[width=0.45\textwidth]{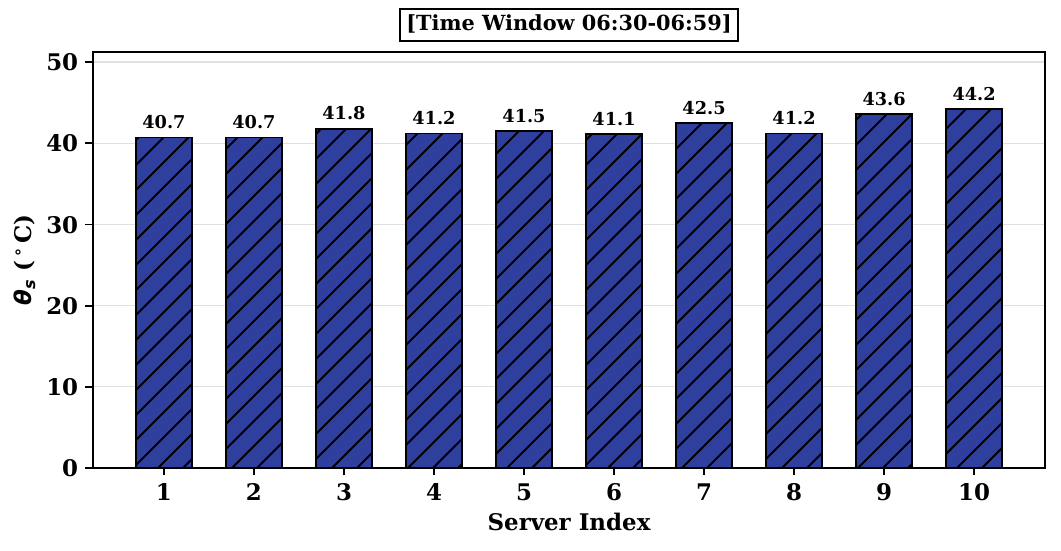}
    \caption{Mean server's exhaust temperature $\theta_{s,i}(t)$ during the first 30-minute interval $[t_\mathrm{06:30},\,t_\mathrm{06:59}]$ after the system scaled to ten active servers. }
    \label{serverapendix1}
\end{figure}

Fig ~\ref{serverapendix1} shows the spatial change in server  temperature across the rack in the 10-server operation window $[t_{06:30},\, t_{06:59}]$. As illustrated in Fig \ref{fig:zonal}, the server close to the cooling unit has smaller index and receives cooler air. Therefore, it exhibits a lower server exhaust temperature. The server temperature difference can be attributed to the relative distance of the server to the cooling unit, resulting in the non-uniform airflow and uneven workload allocation, which may produce more heat in servers with high workload. Collectively, these contributions lead to an increase in server temperatures at the upstream end of the rack.

\subsection{GPU Frequency Sensitivity}
\label{Perclass DVFS}
\begin{figure}[H]
    \centering
    
    \includegraphics[width=0.5\textwidth]{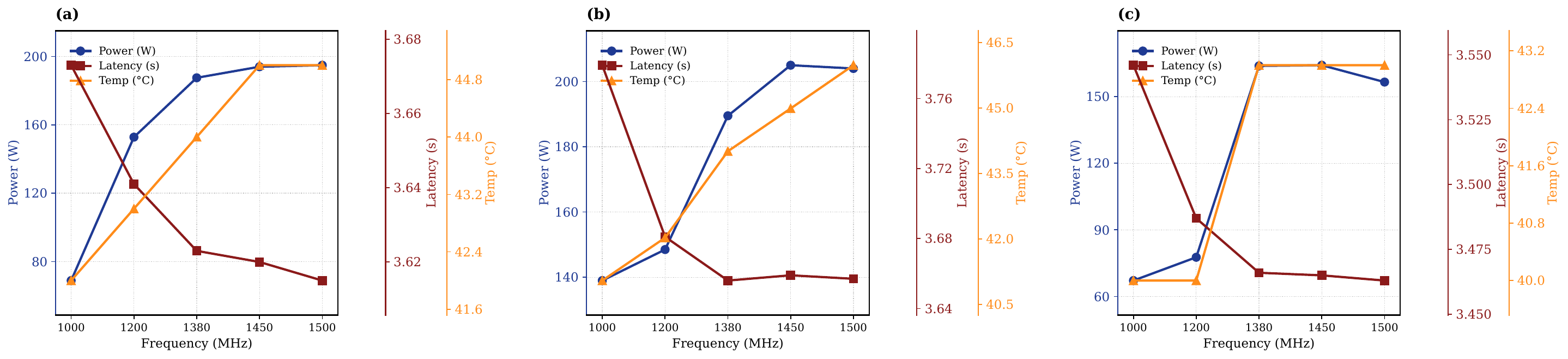}

\caption{GPU frequency sensitivity for \textsc{LLaMA-2-7B} inference on a real 8$\times$Tesla GPU server. We report latency, GPU temperature and  GPU power versus frequency for different token-length classes. Subfigures correspond to token-length classes: (a) Medium, (b) Large, and (c) Small.}
    \label{relationfig7}
\end{figure}

We test on real LLM query data using LlaMA-2-7B model. Fig \ref{relationfig7} illustrates the relationship between GPU frequency, power, latency, and temperature under different token-length categories. It is noteworthy the trained classifier is compatible with such token-length classification. As the frequency of the GPU increases, per-job latency decreases, but power and temperature generally increase. We utilize these profiling data in the DVFS control problem \eqref{eq:DVFS}. This shows the need of class-aware frequency selection to satisfy latency targets while limiting power and temperature.

\subsection{Performance Comparison}

\begin{figure}[H]
    \centering
    
    \includegraphics[width=0.5\textwidth]{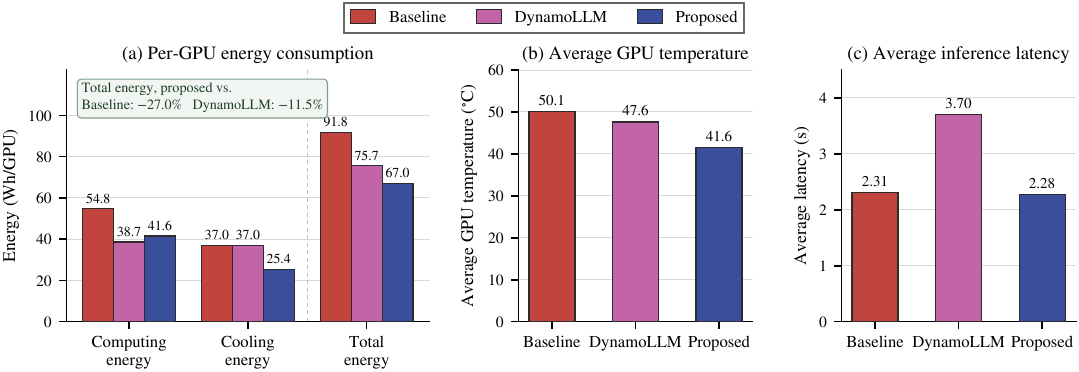}
    \caption{Performance comparison of the baseline, DynamoLLM, and the proposed cooling--computing control framework. (a) Computing and cooling energy consumption per GPU. (b) Average GPU temperature. (c) Average inference latency. Compared with DynamoLLM, the proposed method reduces cooling energy consumption, GPU temperature, and latency, while incurring a modest increase in computing energy. 
}
    \label{figapendix8}
\end{figure}

In Fig.~\ref{figapendix8}, the left panel shows the computing and cooling energy consumption, the middle panel shows the average GPU temperature, and the right panel shows the average inference latency. The results demonstrate that the proposed cooling--computing control framework achieves lower cooling energy consumption, lower GPU temperature, and lower latency than DynamoLLM, highlighting the benefits of jointly optimizing computing and cooling resources.
 
\subsection{Sources of Quantities Used in the Evaluation}
\label{app sources}

In the experimental evaluation, the reported results consist of both directly
measured and model-estimated quantities. Table~\ref{tab:quantity_sources}
summarizes these quantities and clearly distinguishes their respective sources.
In contrast, in the simulation-based evaluation, all reported system quantities
are obtained from the corresponding simulation models.

\begin{table}[H]
\centering
\caption{Sources of quantities used in the evaluation.}
\label{tab:quantity_sources}

\footnotesize
\renewcommand{\arraystretch}{1.05}
\setlength{\tabcolsep}{4pt}

\begin{tabular}{p{0.28\linewidth} p{0.19\linewidth} p{0.43\linewidth}}
\hline
\textbf{Measured} &
\textbf{Trace Input} &
\textbf{Model-Estimated} \\
\hline

GPU power ($P_{\mathrm{GPU}}$) &
Token/job arrivals &
Supply-air temperature ($\theta_{\mathrm{RCU}}$) \\

GPU temperature &
&
Return-air temperature ($\theta_{\mathrm{ret}}$) \\

Latency &
&
Inlet temperature ($\theta_{\mathrm{in}}$) \\

&
&
RCU airflow ($\Phi_{\mathrm{RCU}}$) \\

&
&
Total cooling power ($P_{\mathrm{cooling}}$) \\

\hline
\end{tabular}

\end{table}
\subsection{PID-Based Datacenters Cooling}
\label{PID}
To provide a comparison with the proposed MPC-based cooling controller, we implemented PID controller at the cooling layer. The PID loop uses the tracking error between the measured return-air temperature and its set point to adjust the RCU supply-air temperature command, which is constrained within its physical limits. This setup represents a purely feedback-based, non-predictive controller that is commonly used in practice. For a fair comparison, both MPC and PID controllers were evaluated under same workload and IT power profiles. The results show that the proposed MPC controller reduces the cooling-energy consumption by approximately 28\% while maintaining the return-air temperature closer to its set point than the PID baseline.

\subsection{Model Validation}
\label{ModelVal}
The accuracy of the identified GPU power, GPU temperature and workload forecasting models was evaluated using independent datasets that were not used during parameter estimation. The corresponding validation results are shown in Figs.~\ref{fig}--\ref{fig1}.

\begin{figure}[H]
\centering

\begin{subfigure}[b]{0.48\linewidth}
\centering
\includegraphics[width=\linewidth]{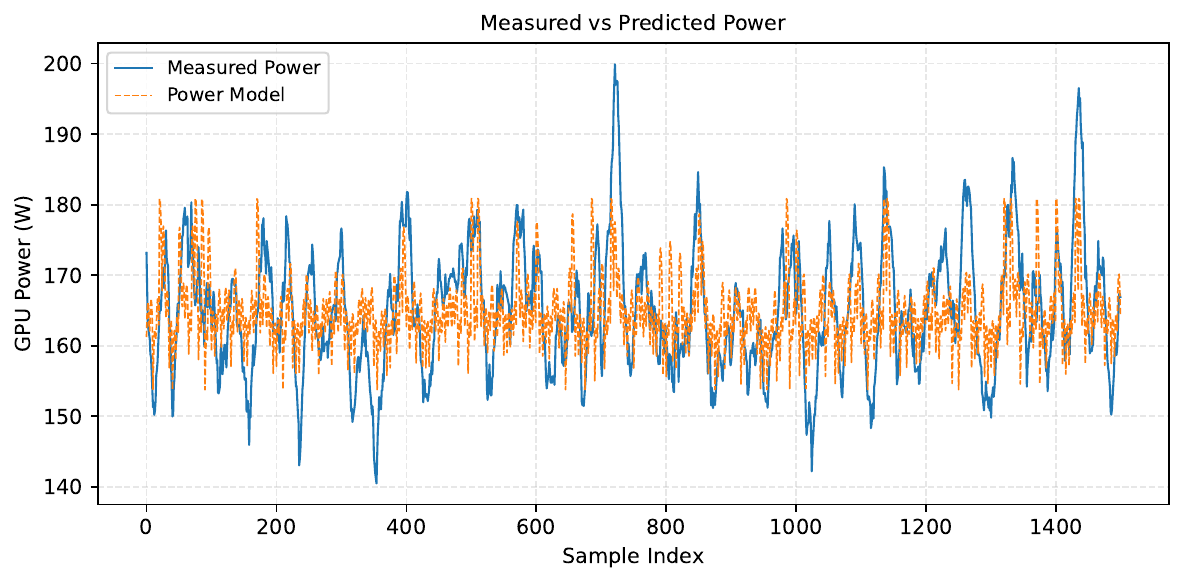}
\caption{GPU power model validation.}
\label{fig}
\end{subfigure}
\hfill
\hfill
\begin{subfigure}[b]{0.48\linewidth}
\centering
\includegraphics[width=\linewidth]{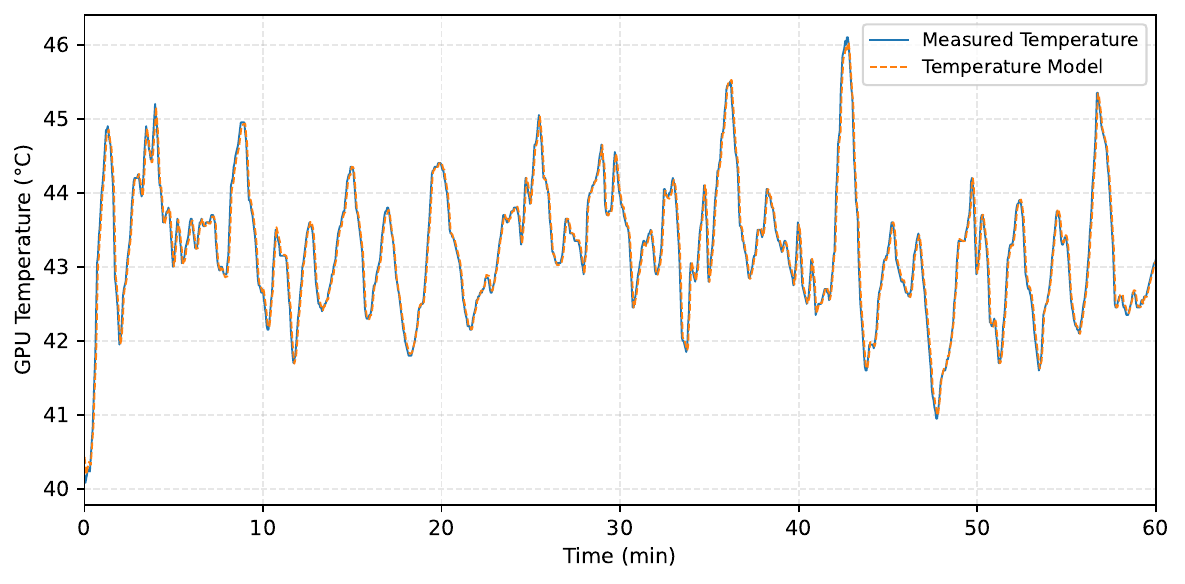}
\caption{GPU temp model validation.}
\label{fig}
\end{subfigure}

\caption{Model validation results using telemetry collected from NVIDIA Tesla V100 GPUs during Llama-2-7B inference experiments. (a) Comparison between measured and predicted GPU power obtained from the identified GPU power model. (b) Comparison between measured and predicted GPU temperature.}
\label{fig}
\end{figure}

The prediction accuracy of the physical models was quantified using the Root Mean Square Error (RMSE) and Mean Absolute Error (MAE). Workload forecasting model was evaluated using the Mean Absolute Percentage Error (MAPE), Normalized Mean Square Error (NMSE) and Normalized Mean Absolute Error (NMAE):
\begin{equation}
\mathrm{MAPE} = \frac{100}{N_L}\sum_{i=1}^{N_L}
\left|\frac{L_i-\hat{L}_i}{L_i}\right|,
\end{equation}

\begin{equation}
\mathrm{NMSE} =
\frac{\sum_{i=1}^{N_L}\bigl(L_i-\hat{L}_i\bigr)^2}
     {\sum_{i=1}^{N_L}L_i^{2}},
\end{equation}

\begin{equation}
\mathrm{NMAE} =
\frac{\sum_{i=1}^{N_L}\bigl|L_i-\hat{L}_i\bigr|}
     {\sum_{i=1}^{N_L}\bigl|L_i\bigr|},
\end{equation}
where $L_i$ and $\hat{L}_i$ denote the actual and predicted token demand, respectively, and $N_L$ is the number of test samples.
\begin{figure}[H]
\centering

\begin{subfigure}[b]{0.7\linewidth}
\centering
\includegraphics[width=\linewidth]{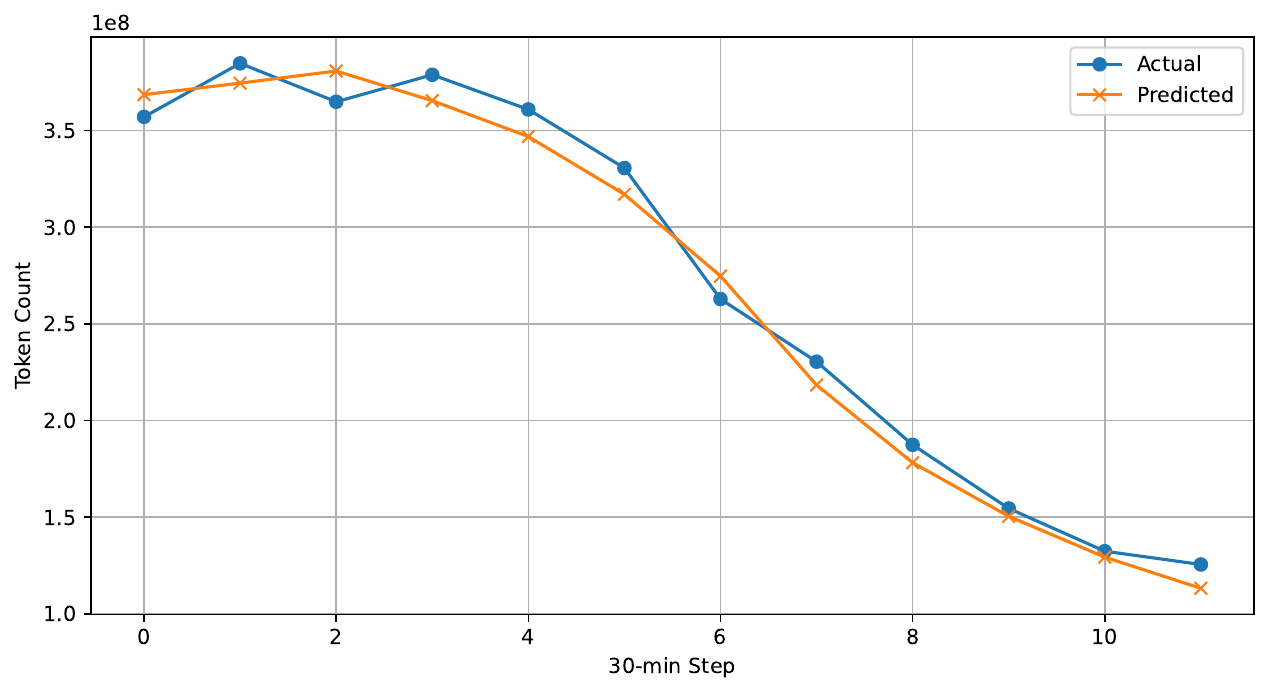}
\caption{workload forecasting validation.}
\label{fig1}
\end{subfigure}
\caption{Validation results for the workload forecasting model. (a) Comparison between actual and predicted token demand using 30-minute prediction intervals.}
\label{fig1}
\end{figure}
Fig.~\ref{fig1}(a) validates the proposed LSTM forecasting model, showing close agreement between the predicted and observed token demand.
\subsection{Physical validation of GPU power and thermal models}
\label{app:model_validation2}

The physical consistency of the identified GPU power and thermal models is
evaluated over the operating range used for model identification. In
particular, we examine the dependence of GPU power on workload utilization
and operating frequency and analyze the steady-state response of the
identified thermal model. Combining the identified power and thermal models also explains the trends
shown in Fig.~\ref{am}(c)--(d). Increasing GPU utilization
or operating frequency increases the predicted GPU power, which in turn
increases the steady-state GPU temperature. Overall, the validation demonstrates
that the identified models produce positive and bounded power and
temperature responses and preserve the expected physical relationships
among workload utilization, GPU frequency, power consumption, and GPU temperature over the considered operating range. The resulting model responses are shown in
Fig.~\ref{am}.
\begin{minipage}[t]{0.9\linewidth}
    \centering
    \includegraphics[width=\linewidth]{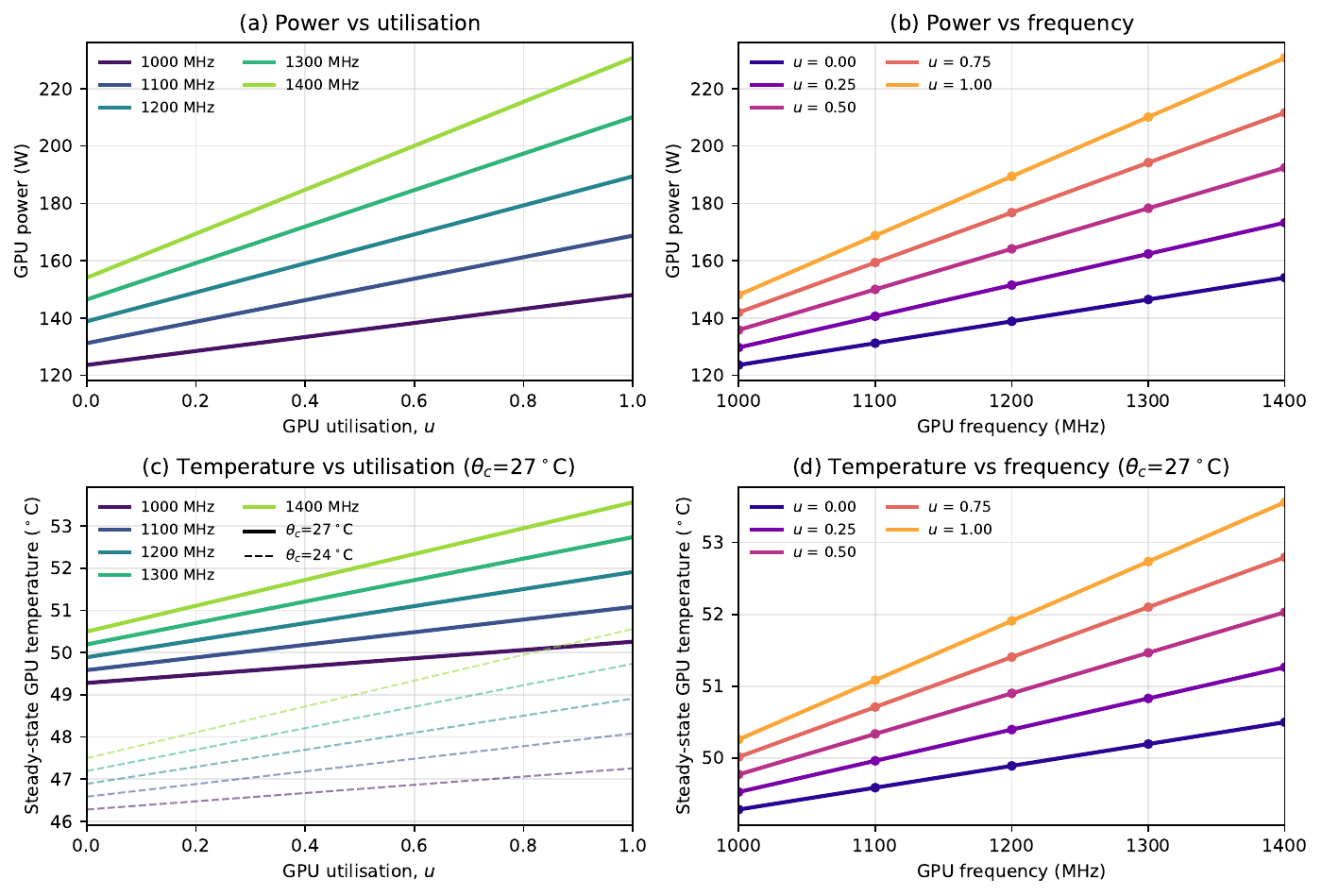}
   
\end{minipage}

\captionof{figure}{Physical validation of the identified GPU power and thermal models
over the considered operating range. (a) GPU power as a function
of utilization at different operating frequencies. (b)  GPU power
as a function of operating frequency at different utilization levels.
(c)steady-state GPU temperature as a function of utilization at
different operating frequencies for cooling inlet temperatures of
$27\,^{\circ}\mathrm{C}$ (solid lines) and $24\,^{\circ}\mathrm{C}$
(dashed lines). (d) steady-state GPU temperature as a  function of
operating frequency at different utilization levels.}
 \label{am}

\subsection{RCU Fan-Power Model Validation}
\label{app:fan_model_validation}

\begin{figure}[H]
    \centering
    \includegraphics[width=0.6\linewidth]{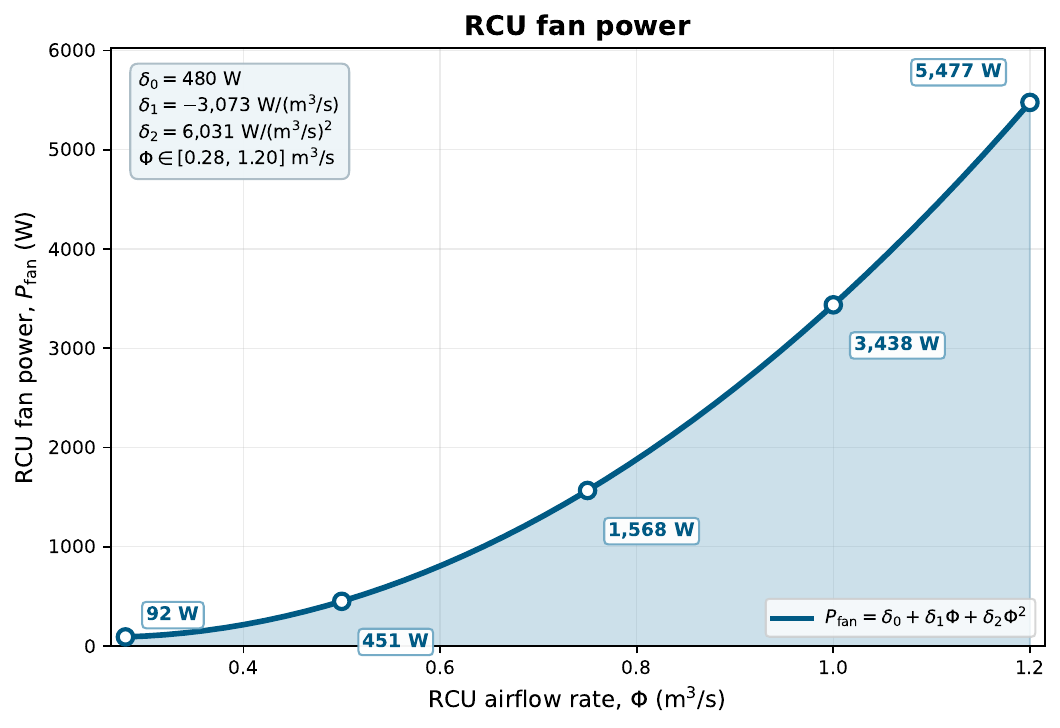}
    \caption{Validation of the adopted rack-based RCU fan-power model over
    the controller operating range. The predicted fan power remains
    positive and increases monotonically as the RCU airflow increases from
    $0.28$ to $1.2$~m$^3$/s.}
    \label{fig:fan_power_validation}
\end{figure}

The RCU fan-power model  is evaluated over
the complete airflow range considered by the cooling controller,
$0.28 \leq \Phi_{\mathrm{RCU}} \leq 1.2$~m$^3$/s. Figure~\ref{fig:fan_power_validation}
shows the resulting fan-power response. Over this operating range, the predicted fan power increases from
approximately $92$~W at $\Phi_{\mathrm{RCU}}=0.28$~m$^3$/s to
$5477$~W at $\Phi_{\mathrm{RCU}}=1.2$~m$^3$/s and remains positive
throughout the entire range. Therefore,
the adopted rack-based RCU fan-power model exhibits a positive and
monotonically increasing relationship between airflow and fan power over
the operating region.

\ifCLASSOPTIONcaptionsoff
  \newpage
\fi